\documentclass{article}

\usepackage{PRIMEarxiv}

\usepackage[utf8]{inputenc} % allow utf-8 input
\usepackage[T1]{fontenc}    % use 8-bit T1 fonts
\usepackage{hyperref}       % hyperlinks
\usepackage{url}            % simple URL typesetting
\usepackage{booktabs}       % professional-quality tables
\usepackage{amsfonts}       % blackboard math symbols
\usepackage{nicefrac}       % compact symbols for 1/2, etc.
\usepackage{microtype}      % microtypography
\usepackage{lipsum}
\usepackage{fancyhdr}       % header
\usepackage{graphicx}       % graphics
\graphicspath{{Figures/}}     % organize your images and other figures under media/ folder

\usepackage{amsmath}
\usepackage{url}
\usepackage{caption}
\usepackage{subcaption}
\usepackage{cleveref}
\usepackage{float}
\usepackage{tabularx}
\usepackage{longtable}
\usepackage{hhline}
\usepackage{adjustbox}

\usepackage[square, numbers, sort&compress]{natbib} 

\usepackage[toc]{appendix}

\usepackage{blkarray}

\usepackage[linesnumbered,ruled,vlined]{algorithm2e}

\usepackage{multirow}

\Crefname{algocf}{Algorithm}{Algorithms}

\DeclareMathOperator*{\argmin}{arg\,min}

\title{Estimation of dynamic Origin-Destination matrices in a railway transportation network integrating ticket sales and passenger count data}

\author{
  Greta Galliani \\
  MOX, Department of Mathematics, Politecnico di Milano, Milan, Italy \\
  \texttt{greta.galliani@mail.polimi.it} \\
   \And
  Piercesare Secchi \\
  MOX, Department of Mathematics, Politecnico di Milano, Milan, Italy \\
  \texttt{piercesare.secchi@polimi.it} \\ 
  \And 
  Francesca Ieva \\
  MOX, Department of Mathematics, Politecnico di Milano, Milan, Italy \\
  HDS, Health Data Science Center, Human Technopole, Milan, Italy \\
  \texttt{francesca.ieva@polimi.it} \\ 
}

\begin{document}
\maketitle

\begin{abstract}
  Accurately estimating Origin-Destination (OD) matrices is a topic of increasing interest for efficient transportation network management and sustainable urban planning. Traditionally, travel surveys have supported this process; however, their availability and comprehensiveness can be limited. Moreover, the recent COVID-19 pandemic has triggered unprecedented shifts in mobility patterns, underscoring the urgency of accurate and dynamic mobility data supporting policies and decisions with data-driven evidence.
  In this study, we tackle these challenges by introducing an innovative pipeline for estimating dynamic OD matrices. The real motivating problem behind this is based on the Trenord railway transportation network in Lombardy, Italy. We apply a novel approach that integrates ticket and subscription sales data with passenger counts obtained from Automated Passenger Counting (APC) systems, making use of the Iterative Proportional Fitting (IPF) algorithm.
  Our work effectively addresses the complexities posed by incomplete and diverse data sources, showcasing the adaptability of our pipeline across various transportation contexts.
  Ultimately, this research bridges the gap between available data sources and the escalating need for precise OD matrices. The proposed pipeline fosters a comprehensive grasp of transportation network dynamics, providing a valuable tool for transportation operators, policymakers, and researchers. Indeed, to highlight the potentiality of dynamic OD matrices, we showcase some methods to perform anomaly detection of mobility trends in the network through such matrices and interpret them in light of events that happened in the last months of 2022. 
\end{abstract}

% keywords can be removed

% SECTIONS OF THE PAPER
\section{Introduction} 
\label{sec:intro} 
The accurate estimation of passenger movements within transportation networks plays a pivotal role in urban planning, infrastructure management and policy formulation. Origin-Destination (OD) matrices, depicting the flow of passengers between various locations within a network, serve as a fundamental tool to capture these travel patterns, providing insights into the dynamics of passenger flows across different regions and modes of transport~\cite{Mohammed_Oke_2022}. Such matrices find applications in diverse fields, from optimizing public transportation services to analyzing the environmental impacts of commuting behaviors~\cite{Mohammed_Oke_2022}.

In recent years, the estimation of OD matrices went through remarkable advancements following the introduction of Automated Data Collection Systems (ADCS) in several transportation networks~\cite{Mohammed_Oke_2022}. However, most works in the field have been anchored in data sources like Automated Fare Collection (AFC) data and travel surveys~\cite{Mohammed_Oke_2022, Cui2006, Gordon2012, Wang2010, Zhao2007, AitAli2019, Torti2021}. These sources provide valuable information for modeling transportation networks but are not always comprehensive or available for all systems. Such limitations pose significant challenges, particularly in complex and multifaceted transportation ecosystems. Moreover, the COVID-19 pandemic has introduced unprecedented shifts in travel behavior, amplifying the need for reliable and up-to-date dynamic mobility data to guide policy responses and recovery strategies~\cite{Hu_2021_2}.

This study takes on the challenge of estimating dynamic OD matrices in the context of a railway transportation network, focusing on the Trenord network operating in Lombardy, Italy. The Trenord network plays a critical role in the mobility landscape of the Lombardy region, connecting urban centers and facilitating daily commutes~\cite{Trenord_chi_siamo}. However, accurate OD matrix estimation for such intricate systems presents obstacles, such as incomplete datasets and the complex interplay between ticket and subscription sales and passenger counts. Indeed, while tickets and subscriptions are sold reporting origins and destinations for the trips, quantitative techniques are needed to assign such trips to specific timeframes and correct for trips happening without purchasing a travel title, accounting for the number of passengers boarding and alighting at each station.

In response to these challenges, we propose an innovative pipeline that combines ticket and subscription sales data and passenger counts collected through Automated Passenger Counting (APC) systems. Our approach addresses data limitations and integrates in a data fusion perspective heterogeneous information sources to deliver accurate dynamic OD matrices that capture the intricate patterns of passenger movement within a portion of the Trenord network.

The main contribution of this work lies in the development of a novel procedure for constructing OD seed matrices from ticket sales data. This multi-step process involves converting ticket and subscription sales records into estimated OD trips, separating trips necessitating transfers, and predicting missing ticket data through gravity models. When combined with passenger count information, the resultant seed matrices undergo iterative refinement using the Iterative Proportional Fitting (IPF)~\cite{IPFP} algorithm to generate dynamic OD matrices. Crucially, our pipeline transcends specific data types, offering a versatile solution applicable to various transportation contexts and adaptable to the desired network's particular characteristics and available data. We show the efficacy of our pipeline by applying it to estimate weekly OD matrices describing trips by train through six train lines of the Trenord network in the period from June to December 2022. Moreover, we showcase tools to perform real-time oversight of complex dynamical systems, aiming to provide methodologies to identify anomalies at global and local station levels in the complex network induced by the dynamic OD matrices. 

The importance of reliable mobility data has been underscored among others by the global pandemic, which has disrupted travel patterns, altered commuter behavior, and prompted shifts in urban dynamics~\cite{Depalma2022}. As cities strive to recover and build resilient transportation systems post-pandemic, accurate mobility insights are indispensable for informed decision-making~\cite{Hu_2021_2}. While extensive work has been dedicated to trip distribution modeling~\cite{Mohammed_Oke_2022}, this study contributes to the present literature by addressing the challenge of utilizing ticket and subscription sales data when AFC data are unavailable. This problem, to the best of our knowledge, has never been approached in past research in the field. We fuse such data with passenger counts derived by the APC system, which has been emerging as the primary tool to estimate passenger counts~\cite{Mohammed_Oke_2022} and whose availability in transportation networks is expected to increase in the future~\cite{Siebert2020}. By doing so, we contribute to a more comprehensive understanding of transportation network dynamics and bridge the gap between available data sources and the demand for accurate, dynamic OD matrices, discussing the implications of our work for transportation management, urban planning, and future research avenues. Through this study, we aspire to contribute to the field of transportation network analysis, offering an adaptable approach for estimating dynamic OD matrices in the intricate context of modern urban mobility. 

The rest of the paper is organized as follows: \Cref{sec:lr} analyzes past literature about the estimation of static and dynamic OD matrices in transportation networks, considering various data sources; \Cref{sec:data} gives some context about the Trenord network and presents the available data to tackle the OD matrices estimation problem; \Cref{sec:methods} recalls some theoretical notions about two well-known techniques in the field of trip distribution modeling, namely gravity models and IPF algorithm and then proceeds to present the pipeline we developed in detail; \Cref{sec:results} shows the results obtained applying said pipeline to estimate weekly OD matrices describing mobility in seven months of 2022 through six train lines of the Trenord network and presents an application of the derived OD matrices to perform anomaly detection in the network; \Cref{sec:discussion} discusses the results obtained, highlighting their strengths and limitations and, finally, \Cref{sec:conclusion} draws the conclusion of our work. In order to enhance reproducibility and transparency, codes are available in the GitHub repository \url{https://github.com/GretaGalliani/dynamic-OD-estimation-railway-network}.

\section{Literature review}
\label{sec:lr}
This Section briefly overviews recent methodologies and approaches for estimating static and dynamic OD matrices, highlighting their strengths, limitations, and recent advancements. Our discussion is grounded in the work of Mohammed and Oke~\cite{Mohammed_Oke_2022}, which provides an extensive literature review of the topic.

The pursuit of estimating OD matrices in transportation networks dates back to the late 1970s~\cite{Wilson1970, Low1972, Robillard1975}. Early efforts heavily relied on surveys and manual data collection, posing challenges due to the sporadic availability of passenger surveys, mainly because of their cost and relevance limitations~\cite{Ben1989}. Recently, advancements in Automatic Data Collection Systems (ADCS) and mobile communication data collection have transformed the landscape~\cite{Mohammed_Oke_2022}. Technologies like Automatic Fare Collection (AFC)~\cite{Cui2006, Gordon2012, Wang2010, Zhao2007, AitAli2019, Torti2021}, Automatic Vehicle Location (AVL)~\cite{Cui2006, Liu2021, Wang2010, Zhao2007}, and Automatic Passenger Counting (APC)~\cite{Cui2006, Liu2021, Torti2021, Wang2010, Zhao2007} have significantly enhanced the accuracy and efficiency of OD estimation. These systems, although not initially designed for integrated OD estimation, offer rich data, reduce collection costs, provide larger sample sizes than surveys, and automate the estimation process \cite{Zhao2004, Cui2006}. As a result, the availability of OD matrices in the future is expected to significantly increase~\cite{Cui2006}.

Among these technologies, AFC systems, particularly smart card data, have become the focal point of recent research~\cite{Cui2006, Gordon2012, Wang2010, Zhao2007, AitAli2019, Wu2021, Torti2021}. Smart cards are widely used in transportation networks across major cities such as Milan, Paris, London, New York, Boston, Beijing, and Hong Kong~\cite{Zannat2019, Torti2021}, offering substantial insights into people's movements. Consequently, smart cards are increasingly replacing surveys in tackling OD estimation challenges. However, estimating accurate OD data for bus or railway networks without AFC systems remains challenging. A potential solution involves leveraging mobile phone data (e.g., Wi-Fi traces, call data records, or global system for mobile communication data) combined with AVL to infer boarding and alighting stops~\cite{Hakegard2018, JafariKang2020, Munizaga2012, Ge2016}. Despite its insights, accessing such data is hampered by privacy concerns~\cite{Hakegard2018}. 

Our work tries to fill this gap, developing an alternative approach to derive OD seed matrices exploiting data readily available to most transportation operators equipped with APC systems but not disposing of AFC ones. Specifically, we leverage ticket and subscription data from a railway network, developing appropriate assumptions to convert this data into OD seeds. These seeds are then combined with passenger counts collected by the APC system to correct the partial estimates accounting for the totality of trips happening in the network. 

In order to tackle this OD estimation challenge, we have at our disposal relevant methodologies that emerged in the field of trip distribution modeling. Among them, growth factor and gravity models are two of the most widely applied techniques~\cite{Mohammed_Oke_2022}. Growth factor models involve updating a seed OD matrix based on growth rates calibrated to match alighting and boarding counts. In this class of models, the Iterative Proportional Fitting (IPF)~\cite{Evans1970, Macgill1977}, also known as the Furness method, is the state-of-the-art approach for estimating OD flows~\cite{Ji2014} and has been employed in several works in the field~\cite{Cui2006, Gordon2012, Liu2021, Zhao2007, Torti2021}. Gravity models, on the other hand, estimate OD matrices using a gravitational attraction model, incorporating population masses and distance or travel cost measures. This model assumes a positive association between flow volume and population size and factors in the effects of distance, space, cost, or travel time on interactions \cite{Wheeler2005}. Other works rely on entropy maximization, proven to be an equivalent approach to gravity models \cite{AitAli2019, Ge2016, Wong2005}. The field of trip distribution modeling also encompasses various other techniques, including maximum likelihood estimation~\cite{Cui2006, Navick1994, Wu2021}, constrained generalized least squares~\cite{Lam2003}, Bayesian estimation~\cite{Hazelton2010, Hakegard2018}, and neural networks~\cite{Mussone2013, Toque2016}.

Lastly, most studies discussed in this review have focused on static OD matrix estimation, describing trips between zones within a fixed period. However, the new abundance of data prompted by the introduction of ACDS prompts interest in dynamic OD matrix estimation, introducing a temporal dimension. Literature in this area is still quite limited~\cite{AitAli2019, Bierlaire2004, Zeng2015, Cerqueira2022} and divides into dynamic a posteriori estimates~\cite{AitAli2019} and real-time estimation for short-term prediction~\cite{Bierlaire2004, Zeng2015}. Our proposed dynamic procedure estimates weekly OD matrices over six months, thus placing itself in the a posteriori estimates field.

\section{Data} 
\label{sec:data}

This Section introduces the Trenord network and the data employed in estimating weekly OD matrices for a specific network segment. 

\subsection{The Trenord network}
Established on May 3, 2011, as a collaboration between FNM and Trenitalia, Trenord is one of the major local rail transport entities in Europe. It boasts an extensive 2,000-kilometer network, interlinking 460 stations, and operates over 2,170 daily trips, serving the Lombardy region in Italy and seven adjoining provinces, along with Malpensa International Airport through the Malpensa Express rail link. The network encompasses 12 suburban lines, 38 regional lines, and 3 lines connecting Lombardy to Switzerland. 77\% of municipalities and 92\% of Lombardy's citizens have a railway station within a 5-kilometer radius. Trenord caters to a substantial daily ridership of more than 550,000 passengers, facilitated by 2,200 train rides~\cite{Trenord_chi_siamo}.

\begin{figure}[htb]
    \centering
    \includegraphics[width=0.5\textwidth]{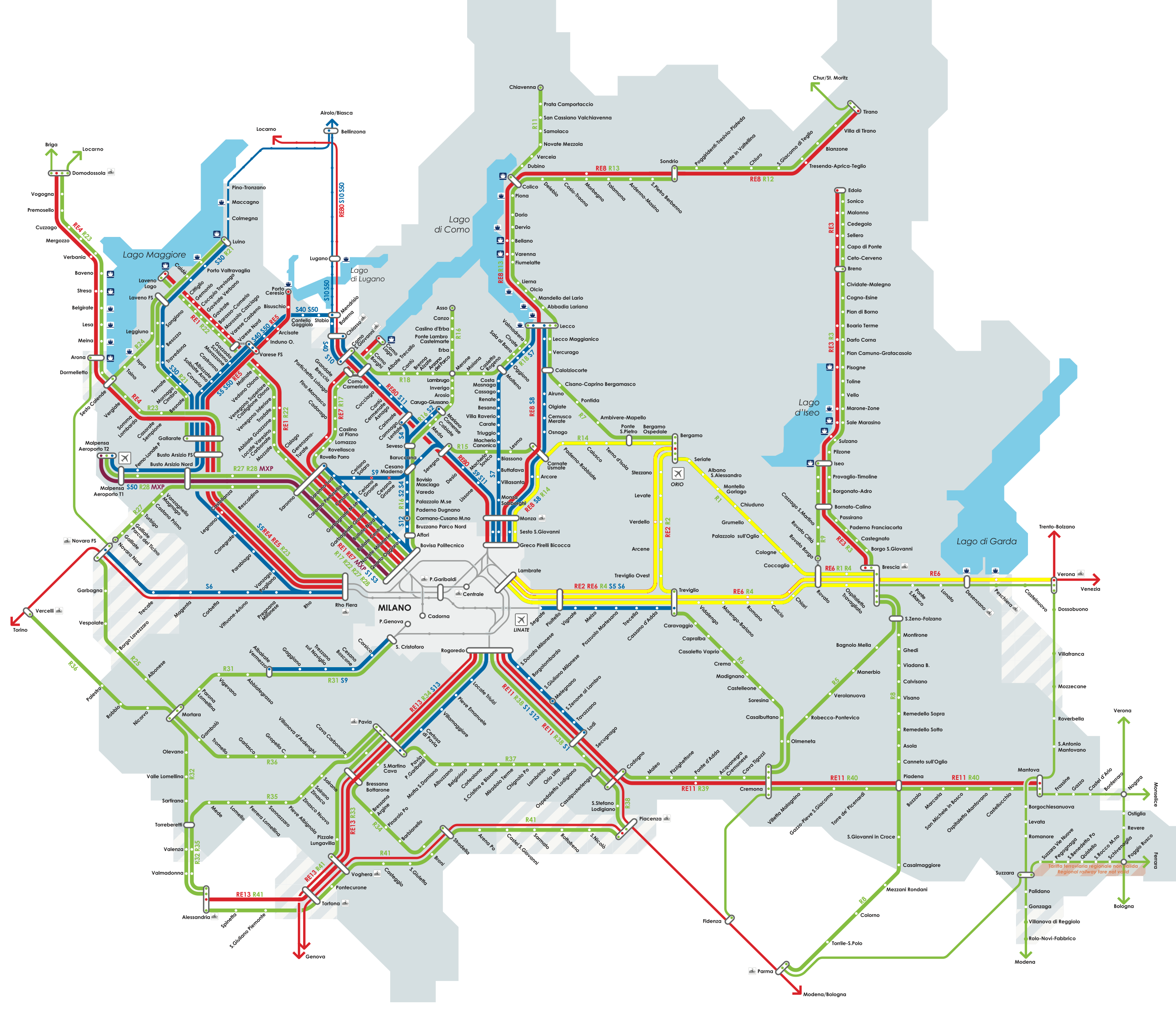}
    \caption[Trenord map of services~\cite{Trenord_map}]{Trenord map of services~\cite{Trenord_map}. Train lines considered in this study are colored in yellow.}
    \label{fig:Trenord_map}
\end{figure}

Our study's scope focuses on six regional lines traversing the provinces of Milan, Brescia, and Bergamo, along with some stations in Lecco and Monza Brianza. Collectively, these lines comprise 46 stations, indicated by the set $S$. \Cref{fig:Trenord_map} visually represents the Trenord network, emphasizing the lines examined in our study. Furthermore, \Cref{table:Trenord_lines} offers an overview of the primary stations of the six train lines under consideration.

\begin{table}[htb]
     \centering
     \begin{tabular}{ll}
       \toprule
       Line     & Main stations   \\
       \midrule
       
       R1 & Bergamo-Brescia \\ 
     R2 & Bergamo-Treviglio \\ 
     R4 & Brescia-Treviglio-Milano \\
     R14 & Bergamo-Carnate-Milano \\
     RE2 & Bergamo-Pioltello-Milano \\ 
     RE6 & Verona-Brescia-Milano \\ 
       \bottomrule
     \end{tabular}
     \\[10pt]
     \caption{Main stations of the Trenord lines involved in the study~\cite{Trenord_lines}}
     \label{table:Trenord_lines}
   \end{table}

\subsection{D\textsuperscript{3}: data sources, data fusion and data engineering}
\label{sec:available_data}
\begin{figure}[htb]
    \centering
    \includegraphics[width=0.5\textwidth]{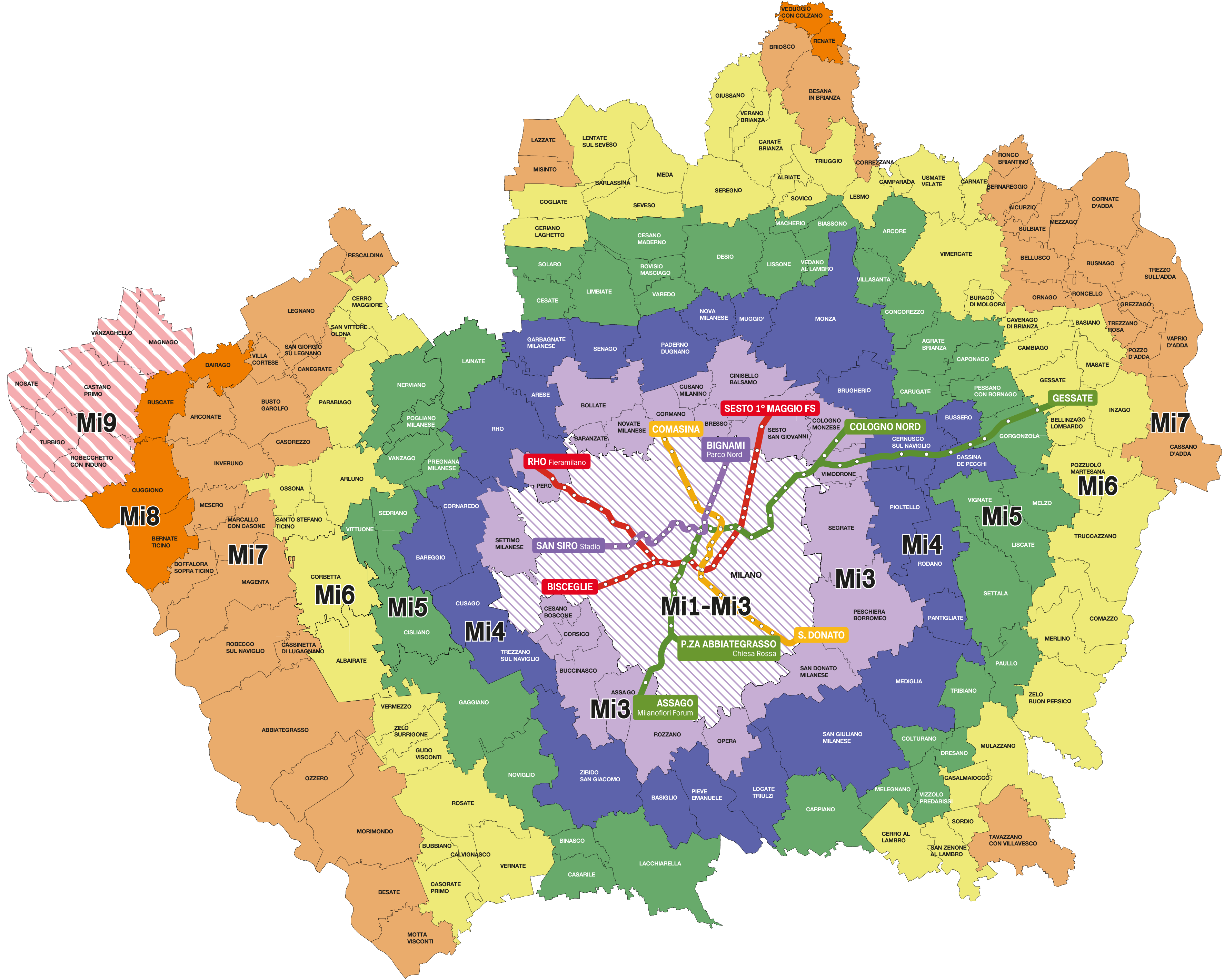}
    \caption{Map of the municipalities belonging to the Integrated Subscriptions area, covering Milan and Monza provinces.}
    \label{fig:IS_area}
  \end{figure}

In addressing the dynamic OD matrices estimation problem, we have at our disposal a range of data provided by Trenord:

\paragraph*{Ticket data} This dataset covers ticket and subscription sales spanning from May 1, 2022, to December 31, 2022, detailing a total of 2,676,629 transactions. The dataset encompasses journeys to and from the 46 stations within our study scope. We will refer to this dataset as the \textit{ticket data}, including regular tickets, carnets, and weekly, monthly, and yearly subscriptions. Each record in the dataset specifies the two stations covered by the ticket, the ticket type, and the date of purchase. Notice that this dataset covers the month of May 2022, which is outside of the estimation period in our study (June-December 2022). The reason is that tickets for trips happening in June can be bought during May. It is essential to acknowledge certain intricacies within the data that necessitate consideration during the estimation process:
\begin{itemize}
    \item Although the ticket data indicates the two stations for which the ticket is issued, information regarding the direction of travel is absent.
    \item The tickets only present the initial and final stations for trips involving transfers, omitting intermediate transfer points.
    \item Tickets encompassing \textit{Verona Porta Nuova} station utilize an interregional fare not documented in Trenord's data, leading to a lack of information regarding Verona-related trips.
    \item Tickets for journeys between Milan stations and stations within the Integrated Subscription (IS) area, illustrated in \Cref{fig:IS_area}, are not accounted for in the sales data. These trips are subject to a distinct fare structure applicable to trips to and from Milan and its environs. 
    \item Tickets sharing fares with another public transport provider operating within Milan and its surroundings are reported at half their actual quantity in the dataset. Since there is no means of deducing the exact number of these tickets sold, we round up the amounts reported in the ticket data to the smallest integers greater than or equal to the said amount.
\end{itemize}

\paragraph*{Counter data} The counter dataset furnishes information regarding passenger boarding and alighting at each station for each train ride belonging to the six study lines. The data is captured through the APC system, deployed on approximately 40\% of the Trenord fleet in 2021~\cite{Trenord_sostenibilita}. The APC system employs sensors on train doors to count passengers, yielding an accuracy error ranging from 5\% to 10\%. In cases where only a train segment is equipped with APC, approximations are employed. When estimation fails, interpolations consider the same train on corresponding days of the preceding weeks, working regressively. In scenarios where reliable estimates cannot be derived, the data lacks passenger counts for the pertinent train rides. Each entry in the dataset provides a status denoting the outcome of the counting process for the respective train ride. The dataset covers the period between June 1, 2022, and December 31, 2022.

\paragraph*{Timetable data} This dataset encompasses departure and arrival times for each station and train ride pertaining to the six study lines. The data spans train rides between June 1, 2022, and December 31, 2022.

\section{Methods} 
\label{sec:methods}

This Section provides an overview of the methods utilized for trip distribution modeling in this study. Two are the main methodologies leveraged in the proposed pipeline: gravity models and the IPF method. Subsequently, we delve into the development of the estimation pipeline designed for this research. 

\subsection{Gravity model for IPF seeds initialization}
\label{sec:gravity_models}
Gravity models have gained widespread use in explaining population movements, commercial trade, and communication patterns~\cite{Simini2012}. Inspired by Newton's law of gravity, the gravity model~\cite{Simini2012, Barbosa2018, Ortuzar2011} suggests that the movement $x_{ij}$ of people, goods, or information between two locations, $i$ and $j$, is influenced by the sizes of these places ($M_i$ and $M_j$, typically related to population or economic scale) and the distance function $f(d_{ij})$ that quantifies their separation in space, time, or cost. Mathematically, this is expressed as:

\begin{equation}
x_{ij} = \mathcal{K} \ \frac{M_i^{\alpha}M_j^{\beta}}{f(d_{ij})}
\label{eq:gravity_model}
\end{equation}

Here, $\alpha$ and $\beta$ represent adjustable exponents, and the distance function $f(d_{ij})$ is often defined using power-law or exponential forms. While the gravity model has gained prominence for approximating travel flows and traffic demand based on local properties, it must be acknowledged that it is a simplified representation and may not fully capture empirical observations~\cite{Simini2012}. Moreover, this model relies on the estimation of multiple parameters, rendering it sensitive to data fluctuations or incompleteness~\cite{Simini2012}.

Our research uses the gravity model to estimate missing data within the ticket-estimated seed OD matrices derived from ticket and subscription sales. This approach enables us to fill missing data with inferred estimates, which are then integrated into the IPF method for merging ticket and counter information.

\subsection{Iterative Proportional Fitting algorithm}
\label{sec:Furness_theory}

Growth-factor models serve as key tools in trip distribution modeling, aiding the adjustment of existing OD matrices using insights into the projected growth of trips originating and terminating within specific zones~\cite{Barbosa2018, Ortuzar2011}. A particularly prominent approach within this category is the IPF method~\cite{Evans1970, Ortuzar2011, Barbosa2018}. This method functions as a doubly constrained growth-factor model, iteratively refining a seed OD matrix to align its row and column sums with data portraying the number of trips originating and ending in each network zone.

In such models, the seed OD matrix can be derived from sources like surveys, historical OD matrices within the transportation network, or insights gleaned from AFC systems, particularly smart card data. In the context of our study, we develop seed OD matrices from ticket and subscription sales data, implementing a structured procedure to translate each ticket type into one or more trips allocated to specific timeframes within the seed OD matrices. Subsequently, we extract data about the volume of passengers boarding and alighting at each station during each timeframe of the study period from counter data collected by the APC system.

Suppose the seed OD matrix $X^*$ takes the form:
\begin{equation*}
\begin{blockarray}{cccc}
\begin{block}{[ccc]c}
 x_{11}^* & \cdots & x_{1J}^* & q_1 \\
 \vdots & & \vdots & \vdots \\
 x_{I1}^* & \cdots & x_{IJ}^* & q_I \\
\end{block}
b_1 & \cdots & b_J & u\\
\end{blockarray}
\end{equation*}

Here, $x_{ij}^*$ denotes the number of trips beginning in zone $i$ and terminating in zone $j$. $I$ represents the count of zones where trips may begin, while $J$ represents the count of zones where trips may end. Further, $q_i = \sum_{j=1}^{J} x_{ij}^*$ signifies the number of trips originating in zone $i$, and $b_j = \sum_{i=1}^{I} x_{ij}^*$ represents the count of trips concluding at zone $j$. The total number of trips is denoted as $u$. Since this matrix does not represent actual network movements, its row and column totals $q_i$ and $b_j$ do not generally match the estimates of trips starting and ending in each zone. Let $p_1, \ldots, p_I$ stand for the estimated count of actual trips originating in each zone, and let $a_1, \ldots, a_J$ denote the estimates for the number of trips ending in each zone. In our application, these estimates are derived from counter data. For consistency, the total number of trips commencing must equal the total number of trips ending, satisfying:

\begin{equation}
 \label{eq:F_consistency}
 \sum_{i=1}^{I} p_i = \sum_{j=1}^{J} a_j = v
\end{equation}

The trip distribution problem is to deduce from matrix $x_{ij}^*$ a forecast matrix $x_{ij}$ whose row and column totals are $p_1, \ldots, p_I$ and $a_1, \ldots, a_J$, respectively:

\begin{equation*}
\begin{blockarray}{cccc}
\begin{block}{[ccc]c}
 x_{11} & \cdots & x_{1J} & p_1 \\
 \vdots & & \vdots & \vdots \\
 x_{I1} & \cdots & x_{IJ} & p_I \\
\end{block}
a_1 & \cdots & a_J & v\\
\end{blockarray}
\end{equation*}

To derive matrix $x_{ij}$, we iteratively determine constants by which to multiply elements within the original matrix $x_{ij}^*$. The IPF method provides a solution to this challenge. In each iteration, we compute matrix $x_{ij}^{(n)}$ by multiplying the previous matrix $x_{ij}^{(n-1)}$ by an appropriate constant $z_{ij}^{(n)}$. In this formulation:

\begin{equation*}
 \begin{aligned}
 & x_{ij}^{(1)} = z_{ij}^{(1)} x_{ij} \\
 & x_{ij}^{(n)} = z_{ij}^{(n)} x_{ij}^{(n-1)} \quad \text{for $n \geq 1$}
 \end{aligned}
\end{equation*}

The matrix $x_{ij}$ represents the limiting matrix $x_{ij} = \lim_{n \to \infty}{x_{ij}^{(n)}}$, and the value $\lim_{n \to \infty} {\prod_{k=1}^{n} z_{ij}^{(k)}}$ serves as the required multiplying factor for $x_{ij}^*$. The detailed iterative procedure is outlined in \Cref{alg:ipf}, with its theoretical foundation in~\cite{Evans1970}.

\begin{algorithm}[htb]
    \DontPrintSemicolon
    \KwIn{Origin and destination marginal totals $p_i$ and $a_j$; Initial trip matrix estimate $x_{ij}^{*}$; maximum number of iteration \texttt{max\_iter}; tolerance \texttt{tol}}
    \KwOut{Estimated trip matrix $x_{ij}$}
    
    Initialize iteration count $k \leftarrow 0$\;
    Initialize convergence flag \texttt{converged} $\leftarrow$ \texttt{False}\;
    Initialize OD matrix $x_{ij}^{(0)} \leftarrow x_{ij}^{*}$\; 
    
    \While{\texttt{not converged}}{
        \For{$i \leftarrow 1$ \KwTo $I$}{
            \For{$j \leftarrow 1$ \KwTo $J$}{
                Compute $x_{ij}^{(k+1)} = x_{ij}^{(k)} \cdot \frac{p_i}{\sum_j x_{ij}^{(k)}} \cdot \frac{a_j}{\sum_i x_{ij}^{(k)}}$\;
            }}

            Check convergence: compute $\epsilon = \max_{i,j}{|x_{ij}^{(k)} - x_{ij}^{(k-1)}|}$\;
            \If{$\epsilon < \texttt{tol}$ || $k \geq max\_iter$}{
                    \texttt{converged} $\leftarrow$ True\;
                }

        Update iteration count: $k \leftarrow k + 1$\;
        }

    Return the estimated trip matrix $x_{ij} = x_{ij}^{(k)}$
    
    \caption{Iterative Proportional Fitting (IPF) Algorithm}
    \label{alg:ipf}
\end{algorithm}

After computing matrix $X$, the problem of evaluating its goodness of fit arises. Since we have no data available describing the ground truth of railway movements $x_{ij}^{[w]}$, we provide an evaluation through the maximum deviation between the generated and desired margin, expressed as:
\begin{equation}
\label{eq:Furness_errors}
    \begin{aligned}
        \epsilon_{row} = \max_i \left\lvert p_i - \sum_{j=1}^{J} x_{ij}  \right\rvert \\
        \epsilon_{col} = \max_j \left\lvert a_j - \sum_{i=1}^{I} x_{ij}  \right\rvert 
    \end{aligned}
\end{equation}

During our application, we encountered specific challenges inherent to the IPF method. These challenges and their potential solutions are thoroughly discussed in~\cite{IPFP_problems}. We discuss the solutions we adopted to three of these problems:
\begin{enumerate}
    \item \textbf{Zero cell problem:} 
    The IPF method can not correct zero cells directly. Hence, it is necessary to identify and address cells that require modification by the algorithm. Although zero cells often signify the impossibility of travel between two zones, they can also stem from erroneous estimates during surveys. In our approach, we tackle this issue by substituting zero cells in the seed matrices with a value of 0.1. This substitution applies exclusively to cells describing direct paths $(i,j) \in \mathcal{D}$, i.e., those connecting two stations both belonging to one of the six train lines, while cells corresponding to indirect routes remain at zero. This approach aligns with our core objective, which is to estimate the number of train trips between each pair of stations, thus differentiating between trips requiring transfers. Subsequently, undirect path (i.e. paths requiring at least one transfer) should have zero cells in the resulting OD matrices, $x_{ij} = 0 \ \forall (i,j) \notin \mathcal{D}$.
    \item \textbf{Marginals consistency:} 
    While Equation~\ref{eq:F_consistency} stipulates that the total number of trips starting should equal the total number of trips ending, this assumption is disrupted in our context due to estimation errors in marginals $p_1, \ldots, p_I$ and $a_1, \ldots, a_J$. To overcome this inconsistency, \cite{IPFP} proposes a solution involving a shift toward probabilities. This entails defining the following:
    \begin{equation}
     \label{eq:Furness_probability}
       \begin{aligned}
            \pi_{ij}^* = \frac{x_{ij}^*}{\sum_{ij}{x_{ij}^*}} \\
            \rho_i = \frac{p_i}{\sum_{i}{p_i}} \\
            \alpha_j = \frac{a_j}{\sum_{j}{a_j}}
       \end{aligned}
   \end{equation}
   By performing this operation, consistency is restored, as $\sum_{i=1}^{I} \rho_i = \sum_{j=1}^{J} \alpha_j = 1$. Subsequently, the IPF method is employed using $\pi_{ij}^*$ as the seed matrix and $\rho_1, \ldots, \rho_I$ and $\alpha_1, \ldots, \alpha_J$ as marginals. This results in the matrix $\pi_{ij}$ where $\sum_{ij} \pi_{ij} = 1$. Each cell within $\pi_{ij}$ represents the probability of a single trip occurring from zone $i$ to zone $j$. To deduce the OD matrix estimating the actual number of trips between each zone pair, we must multiply matrix $\pi_{ij}$ by the total number of trips, $v$. The selection of $v$ can be either the total number of boarded passengers ($\sum_{i=1}^{I} p_i$) or the total number of alighted ones ($\sum_{j=1}^{J} a_j$).
   \item \textbf{Integer conversion:} 
   The matrix $x_{ij}$ entries denote the number of trips originating from zone $i$ and terminating in zone $j$. Therefore, these entries should naturally be integer values. However, the IPF method generates a matrix with non-integer values. To address this, we apply integer conversion to the output matrix. This approach has drawbacks, particularly concerning information discrimination, implying that cells containing values such as 0.501 and 0.999 are treated equally after conversion~\cite{IPFP_problems}. Moreover, the question of whether the seed $x_{ij}^*$ and marginals $p_i$ and $a_j$ should be integers arises. These values hold natural interpretations in terms of estimated trips ($x_{ij}^*$), the number of trips originating in zone $i$ ($p_i$), and the number of trips ending in zone $j$ ($a_j$). We opt to round the marginal vectors $p_i$ and $a_j$, while rounding is not applied to the seed OD matrices $x_{ij}^*$ cells, as fractional values could result from the procedure used to generate these seeds described in~\Cref{sec:pipeline_ticket}.
\end{enumerate}

\subsection{Estimation pipeline}
\label{sec:pipeline_theory}
\begin{figure}[htb]
    \centering
    \includegraphics[width=0.8\textwidth]{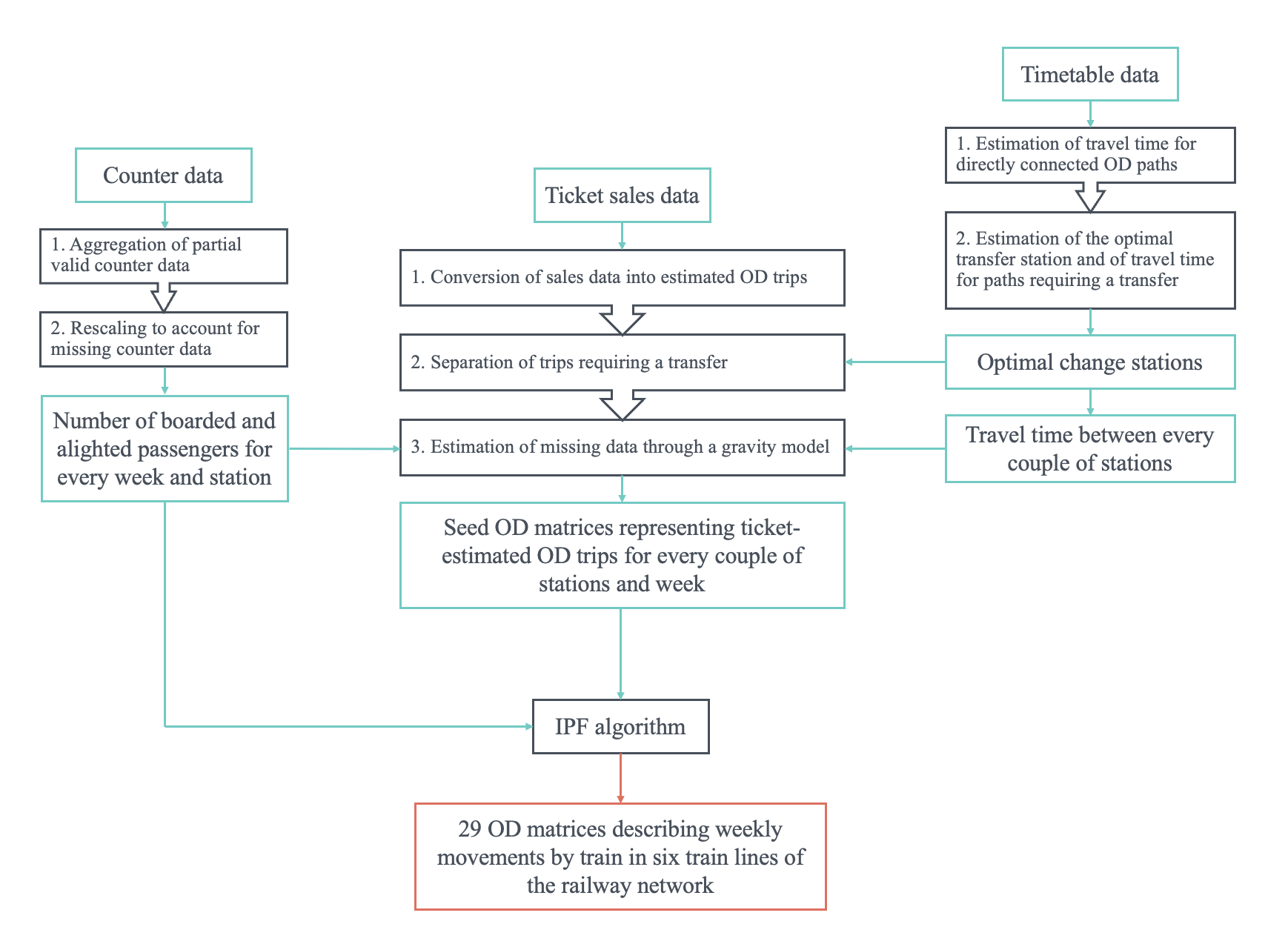}
    \caption{Estimation pipeline developed to obtain dynamic OD matrices describing movements in the Trenord network for a given time period divided into timeframes, using ticket, timetable, and counter data. Green boxes represent input and processed data, arrows indicate the data flow through the pipeline, black boxes denote the pipeline's steps, and the red box highlights the final pipeline's results.}
    \label{fig:pipeline}
   \end{figure}

In this Section, we outline the pipeline used to estimate weekly OD matrices describing train movements across stations $i \in S$ of the Trenord network for a set $w \in W$ of timeframes individuated in a time period. The input data are the ticket, counter, and timetable datasets described in~\Cref{sec:available_data}. The pipeline progresses through the series of steps shown in~\Cref{fig:pipeline}.

\subsubsection{Estimation of mean travel time and optimal change stations}
\label{sec:pipeline_times}
Starting with the timetable data, which describes the actual departure and arrival times of each train ride in the considered portion of the Trenord network, we compute the mean travel time for each direct path, i.e., those connecting two stations belonging to one of the train lines covered by data. To ensure data quality, we eliminate values outside the [0.05, 0.95] quantile range that could stem from data entry errors or extreme delays. The computed matrix $t_{ij}$ stores mean travel times for each direct path $(i,j) \in \mathcal{D}$, with the fastest travel time chosen for stations connected by multiple lines. Notice that while we estimate travel times from actual timetable data, these times could alternatively be collected from the service's schedule. 

This data is then used to estimate the optimal change station for paths not directly connected. For each route $(i,j) \notin \mathcal{D}$ without a direct connection, we identify the optimal transfer station $k^*$ as follows:
\begin{equation*}
 k^* = \argmin_k {t_{ik} + t_{kj}} \ \forall k : \exists (i,k),(k,j) \in \mathcal{D}
\end{equation*}

The travel time between non-directly connected stations is estimated as follows:
\begin{equation*}
 t_{ij} = t_{ik^*} + t_{k^*j}
\end{equation*}
Thus, the matrix $t_{ij}$ includes travel times for directly connected stations and those that can be connected with at most one transfer. Notably, travel times involving multiple transfers are omitted since we evaluated that estimating more than one transfer station would introduce significant uncertainty in the procedure.  Indeed, the choice of the transfer station also depends on the transfer waiting time, which we did not account for in the computation of $t_{ij}$ since waiting times depend on the specific choice of the train rides and the time of the day when the trip happens. 

\subsubsection{Estimation of boarded and alighted passengers from counter data}
\label{sec:pipeline_counter}
Accurate estimates of boarded and alighted passengers for each station $i \in S$ during each timeframe $w \in W$ are obtained using the counter data collected by the APC system. For each station-timeframe combination $(i,w)$, we compute partial estimates of boarded and alighted passengers ($partial \_ boarded_i^{[w]}$ and $partial \_ alighted_i^{[w]}$) based on valid counter data, aggregating data collected by the APC system or estimated through interpolation. The coverage of each station-timeframe pair, $coverage_i^{[w]}$, is calculated as the ratio of train rides with valid counter data to the total train rides in timeframe $w$ stopping at station $i$:
\begin{equation*}
    \label{eq:Coverage}
    coverage_i^{[w]} = \frac{\#\{train \ rides \ having \ valid \ counter \ data \}_i^{[w]}}{\#\{total \ train \ rides \}_i^{[w]}}
\end{equation*}

We can estimate the total boarded and alighted passengers by computing the mean number of boarded and alighted passengers in the trains having valid counter data and then multiplying this mean for the number of trains stopping in station $i$ during week $w$. This procedure is equivalent to the following estimators:
\begin{equation}
    \label{eq:Rescaling_trains}
    \begin{aligned}
        & p_i^{[w]} = \frac{partial \_ boarded_i^{[w]}}{\#\{train \ rides \ having \ valid \ counter \ data \}_i^{[w]}} \ *\ \#\{total \ train \ rides \}_i^{[w]}\\
        & a_i^{[w]} = \frac{partial \_ alighted_i^{[w]}}{\#\{train \ rides \ having \ valid \ counter \ data \}_i^{[w]}} \ *\ \#\{total \ train \ rides \}_i^{[w]}\\
    \end{aligned}
\end{equation}

Alternatively, they can be expressed as:
\begin{equation*}
    \begin{aligned}
        & p_i^{[w]} = \frac{partial \_ boarded_i^{[w]}}{coverage_i^{[w]}} \\
        & a_i^{[w]} = \frac{partial \_ alighted_i^{[w]}}{coverage_i^{[w]}} \\
    \end{aligned}
\end{equation*}

This step assumes adequate coverage of counter data for a significant fraction of train rides in the network. If this assumption is not respected, alternative methods to estimate the passenger counts should be discussed, such as regression models considering predictors like the number of trains stopping at a station during the week, the number of ticket-estimated incoming and outgoing trips, and other relevant factors. Moreover, while not explored in this work, the rescaling procedure could be stratified to account for weekdays and timeslots of train rides, aiming to obtain passenger count estimates closely resembling reality. 

\subsubsection{Conversion of ticket and subscription sales records into Origin-Destination seeds}
\label{sec:pipeline_ticket}
This step generates seed OD matrices from ticket and subscription sales to be iteratively refined using the IPF algorithm. We establish assumptions linking each record in the ticket dataset to estimated trips, suggested by the data provider considering information on ticket validity and usage~\cite{Trenord_tickets}. The premises for attributing ticket records to cells in the dynamic seed OD matrices $x_{ij}^{[w]*}$ are detailed in~\Cref{appendix:AppendixA}. This step forms the cornerstone for producing OD seeds. Assumptions may evolve based on changes in ticket regulations or other factors, and future work could involve robust sensitivity analysis, exploring the impact of altering these assumptions on the final OD estimates.

Applying these assumptions, we derive a number of $\#{W}$ OD matrices representing estimated trips for all possible OD paths. Notice that paths not directly connected by a single train line are covered by tickets reporting only the initial and final stations. We distinguish such routes by leveraging the optimal transfer stations obtained in~\Cref{sec:pipeline_times}. For each OD path $(i,j)$ requiring at most one transfer, we decompose the ticket-estimated OD entries $x_{ij}^{[w]*}$ into $x_{ik^*}^{[w]*}$ and $x_{k^*j}^{[w]*}$, where $k^*$ is the optimal transfer station. This separation emphasizes our goal of estimating dynamic OD matrices for train trips in the Trenord network, treating trips requiring transfers as two distinct journeys. However, trips needing more than one transfer are excluded due to the associated uncertainty, as discussed in~\Cref{sec:pipeline_times}. As a result, cells of the matrices $x_{ij}^{[w]*} = 0$ where $(i,j) \notin \mathcal{D}$.

To handle some routes with missing ticket sales data, we employ a gravity model to predict missing values, ensuring reliable seed matrices, which are critical for accurate OD matrix estimation through the IPF method~\cite{IPFP_problems}. By log-transforming the gravity model of~\Cref{eq:gravity_model}, we can estimate the parameters through Ordinary Least Squares, predicting the missing paths for each timeframe of the study $w \in W$. The log-transformed gravity model is mathematically formulated as follows:
\begin{equation}
    \label{eq:gravity_model_Trenord}
    log(x_{ij}^{[w]*}) = log(\mathcal{K}) + \alpha \ log(p_i^{[w]}) + \beta \ log(a_j^{[w]}) + \gamma \ log(t_{ij}) + \epsilon_{ij}
\end{equation} 
The masses are chosen to be the number of passengers boarding for the origin station $p_i^{[w]}$ and alighting at the destination one $a_j^{[w]}$ in the considered timeframe $w \in W$, while the distance function is a power law of the travel time estimated in~\Cref{sec:pipeline_times}. For cells $x_{ij}^{[w]*}=0$, we substitute $x_{ij}^{[w]*} = 0.01$ to fit the model. We also exclude stations with $p_i^{[w]} =0$ or $a_j^{[w]}=0$, since the natural entry for these cells is $x_{ij}^{[w]*}=0$. A linear regression model is then used to predict the missing paths for each timeframe $w \in W$.

Lastly, when making predictions, some predicted cells are not empty before the filling since the separation of OD paths requiring a train change may have added trips to the paths. In these cases, we discard this information and fill the cells uniquely by the prediction of the linear regression models since it is expected to account for all the trips, including those involving transfers. 

We chose to not account for trips involving more than one transfer, as these constitute a minority of the ticket data. Indeed, we estimate that this kind of trip contitutes less than 1\% of the total number of estimated trips, as shown in~\Cref{table:tickets_transfers},~\Cref{sec:results}. The estimation of trips with multiple transfers could be explored in future research, especially for broader networks. Additionally, the deterministic approach employed to identify optimal transfer stations could be enhanced to consider waiting times and potential passenger preferences. For instance, the change station for each trip through stations not connected by a train line could be extracted in a set of candidates considering a probability inversely proportional to the total travel time.

\subsubsection{Summary of the Iterative Proportional Fitting method}
\label{sec:pipeline_IPF}
To obtain dynamic OD matrices describing trip counts $x_{ij}^{[w]}$ for each timeframe $w \in W$, we apply the IPF algorithm described in~\Cref{sec:Furness_theory} individually to each of the $w$ timeframes. The ticket-estimated OD matrices $X^{*[w]}$ serve as seed matrices, with boarded passengers $p_i^{[w]}$ and alighted passengers $a_j^{[w]}$ as margins.

Furthermore, to tackle the zero correction problem explained in~\Cref{sec:Furness_theory}, where some $x_{ij}^{*[w]}$ cells might be zero, we replace such cells corresponding to direct paths $(i,j) \in \mathcal{D}$ with the artificial value 0.1, while we leave cells corresponding to indirect routes to 0.

Due to estimation errors and APC system issues, $\sum_{i=1}^{I}{p_i^{[w]}} \neq \sum_{j=1}^{J}{a_j^{[w]}}$ for any given timeframe $w \in W$, contradicting the consistency assumption in~\Cref{eq:F_consistency}. To address this, we scale tickets, boarded, and alighted passengers using the probability interpretation defined by~\Cref{eq:Furness_probability}, yielding:

\begin{equation*}
    \label{eq:scaled_passengers}
\begin{aligned}
    \pi_{ij}^{*[w]} = \frac{t_{ij}^{*[w]}}{\sum_{(i,j) \in S}{x_{ij}^{*[w]}}} \\
    \rho_i^{[w]} = \frac{p_i^{[w]}}{\sum_{k \in S}{p_k^{[w]}}} \\
    \alpha_j^{[w]} = \frac{a_j^{[w]}}{\sum_{k \in S}{a_k^{[w]}}} \\
\end{aligned}
\end{equation*}

This adjustment ensures $\sum_{i \in S}{\rho_i^{[w]}} = \sum_{j \in S}{\alpha_j^{[w]}} = 1 \ \forall w \in W$ and $ \sum_{(i,j) \in S}{\pi_{ij}^{*[w]}} = 1 \ \forall w \in W$.

Applying the IPF method results in the matrix $\pi_{ij}^{[w]}$, which is then corrected by multiplying each cell by the total number of boarded passengers as:
\begin{equation*}
    x_{ij}^{[w]} =  \pi_{ij}^{[w]} \sum_{k \in S}{p_k^{[w]}}
\end{equation*} 
This adjustment ensures the final matrix $X^{[w]}$ accurately represents the trips originating in station $i$ and ending in station $j$ during timeframe $w$. Alternatively, the total number of alighted passengers in week $w$, $\sum_{k \in S}{a_k^{[w]}}$, could be used for scaling.

After this procedure, the final matrix $X^{[w]}$ is rounded, as its entries $x_{ij}^{[w]}$ naturally represent the number of trips originating in station $i$ and ending in station $j$ during timeframe $w$.

\section{Results} 
\label{sec:results}
In this Section, we present the results obtained by applying the proposed pipeline to estimate weekly OD matrices that describe train movements across six train lines within the Trenord network for each week in the seven-month period of June-December 2022. Thus, the set of stations $S$ covered by the estimation pipeline presented in~\Cref{sec:pipeline_theory} comprises 46 stations of the Trenord network and the set of timeframes $W$ consists of the 29 weeks from June 6, 2022, to December 25, 2022. We conducted statistical analyses using the R software environment~\cite{R} and employed the \texttt{mipfp} package to execute the IPF algorithm~\cite{IPFP}.

\subsection{Exploratory analysis}
\begin{figure}[htb]
    \centering
    \includegraphics[width=0.7\textwidth]{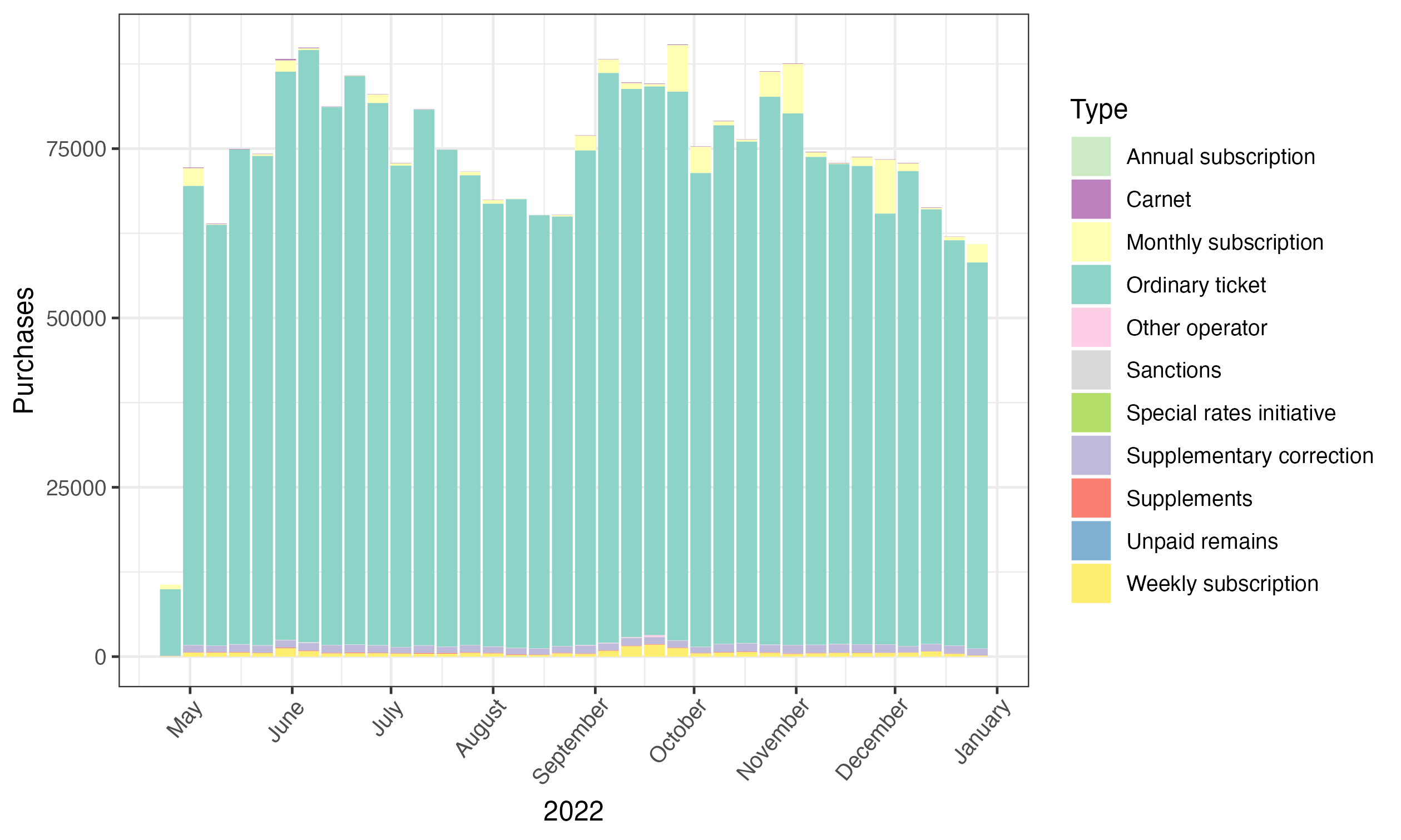}
    \caption{Number of purchased tickets and subscriptions across seven months of 2022, stratified by type.}
    \label{fig:tickets}
\end{figure}

We first present an exploratory analysis conducted on the datasets provided by Trenord, described in~\Cref{sec:available_data}, to highlight some criticalities in the data. \Cref{fig:tickets} displays the number of tickets purchased during the study period, distinguishing ticket types. \Cref{table:tickets} shows the sales for each ticket type throughout the seven months of 2022 considered in the study. We can notice that ordinary tickets are by far the most purchased kind of ticket. Moreover, we observe a reduction in the purchases of all types of tickets in the summer period (July and August), followed by an increase in September after the end of the summer holidays. 

\begin{table}[htb]
    \centering
    \begin{tabular}{lll}
      \toprule
      Ticket type     & Purchases & \% of total ticket sales data   \\
      \midrule
      Ordinary tickets & 2,559,799 & 95.6\% \\
      Monthly subscriptions & 51,076 & 1.91\% \\ 
      Weekly subscriptions & 21,106 & 0.79\% \\
      Carnets & 2,476 & 0.09\% \\ 
      Annual subscriptions & 194 & 0.01\% \\ 
      Other operator & 1,172 & 0.04\% \\
      Sanctions & 1,012 & 0.04\% \\
      Special rates initiatives & 6  & <0.01\% \\
      Supplementary corrections & 37,029 & 1.38\% \\
      Supplements & 2,625 & 0.1\% \\  
      Unpaid remains & 134 & 0.01\% \\

      \bottomrule
    \end{tabular}
    \\[10pt]
    \caption{Number of purchases and percentual of total sales for each ticket type during the seven months of 2022 considered in the study.}
    \label{table:tickets}
  \end{table}

\begin{figure}[htb]
    \centering
    \includegraphics[width=0.7\textwidth]{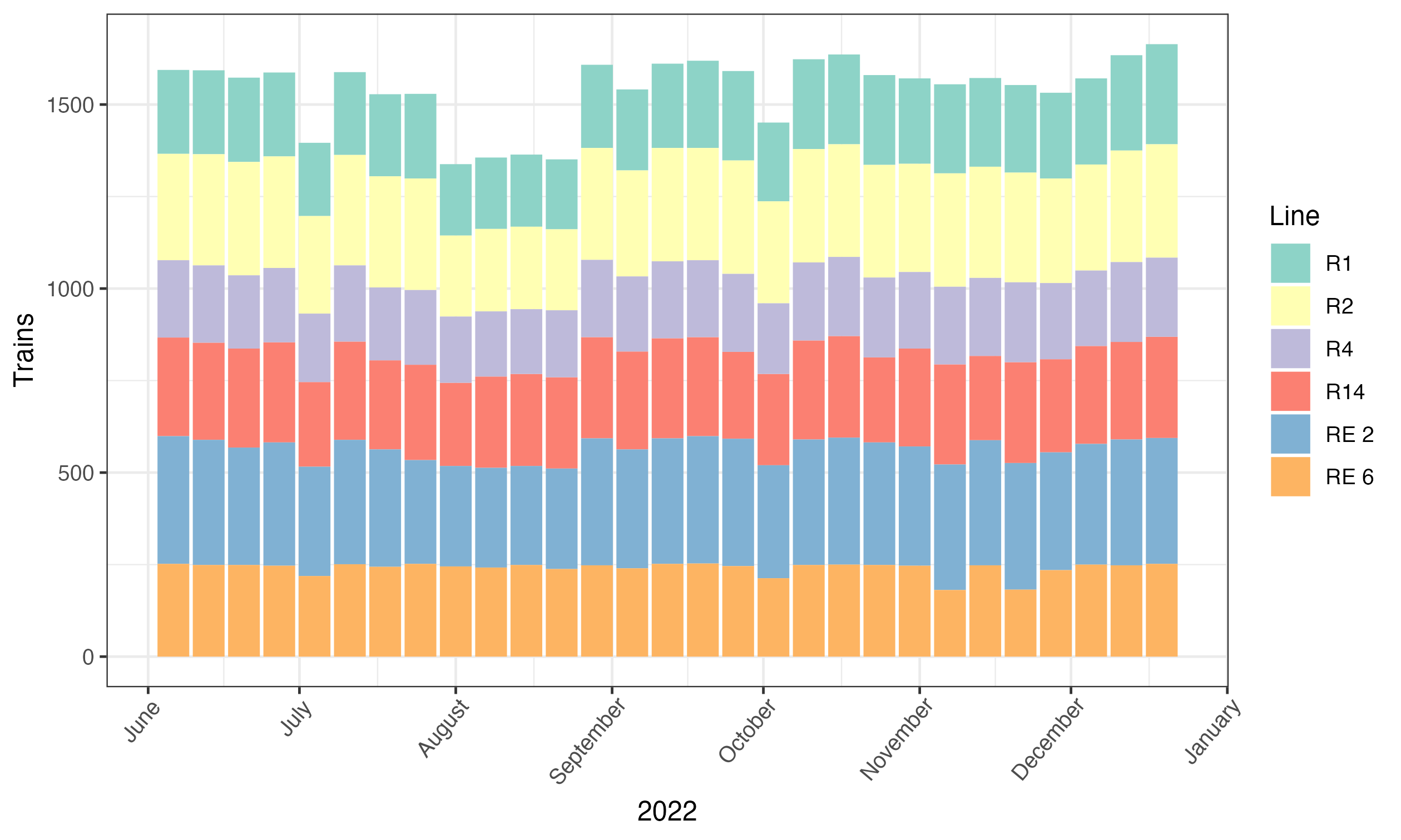}
    \caption{Number of trains per week during seven months of 2022, stratified by the six train lines.}
    \label{fig:trains}
\end{figure}

Turning to the data collected by train counters,~\Cref{fig:trains} shows the number of trains analyzed in our study. Once again, we note a slight reduction in train rides' number during August.

\begin{figure}[htb]
    \centering
    \includegraphics[width=0.7\textwidth]{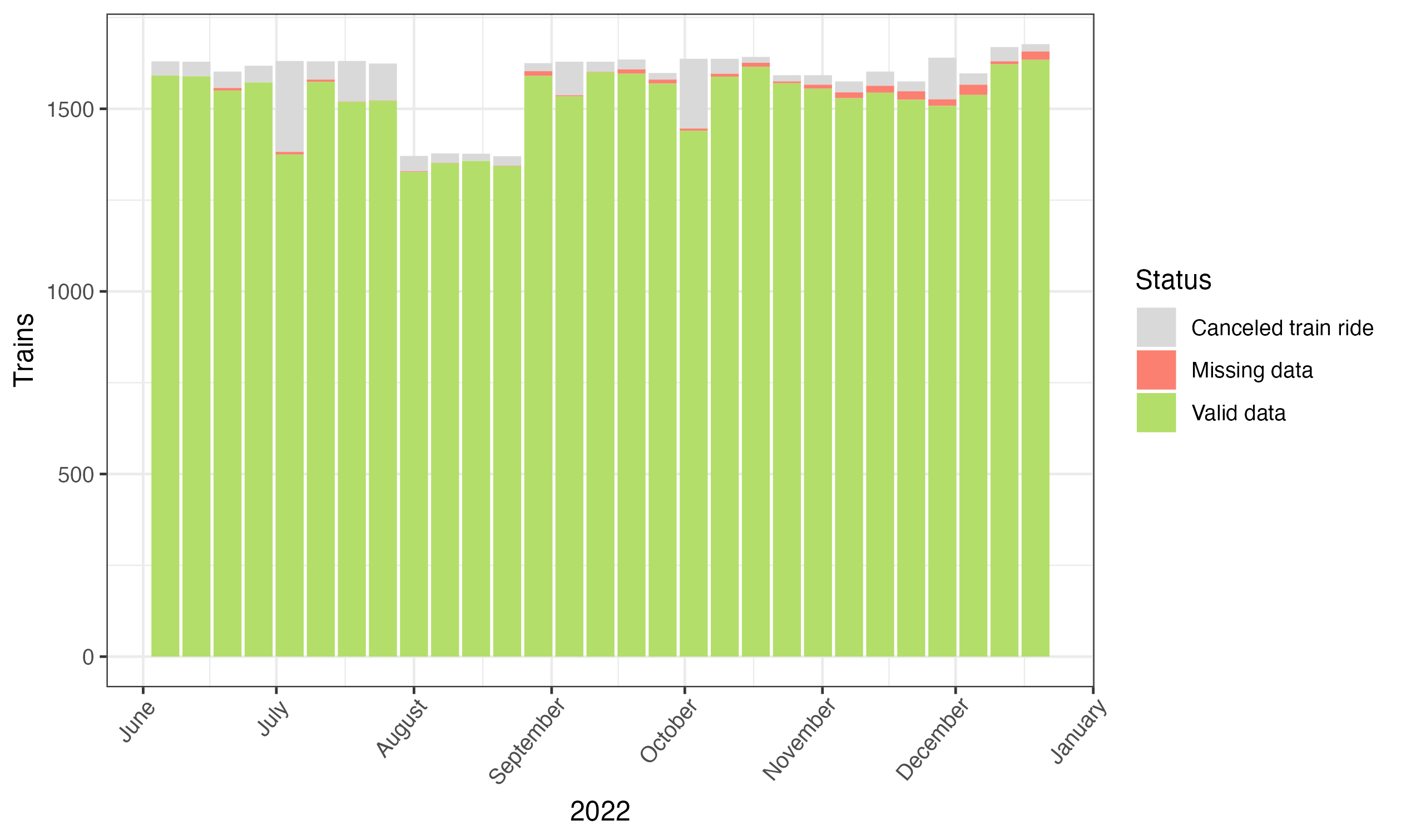}
    \caption{Distribution of the APC system's states of train rides during seven months of 2022.}
    \label{fig:APC_states}
\end{figure}

Finally,~\Cref{fig:APC_states} illustrates the distribution of the states of the APC system within train rides. The states include valid data, missing data, and canceled train rides (where no passengers could board or alight the train). Missing data accounts for only 0.5\% of train rides, excluding the canceled ones.

\subsection{Counter data aggregation and estimation of missing data}
\begin{figure}[htb]
    \centering
    \includegraphics[width=0.7\textwidth]{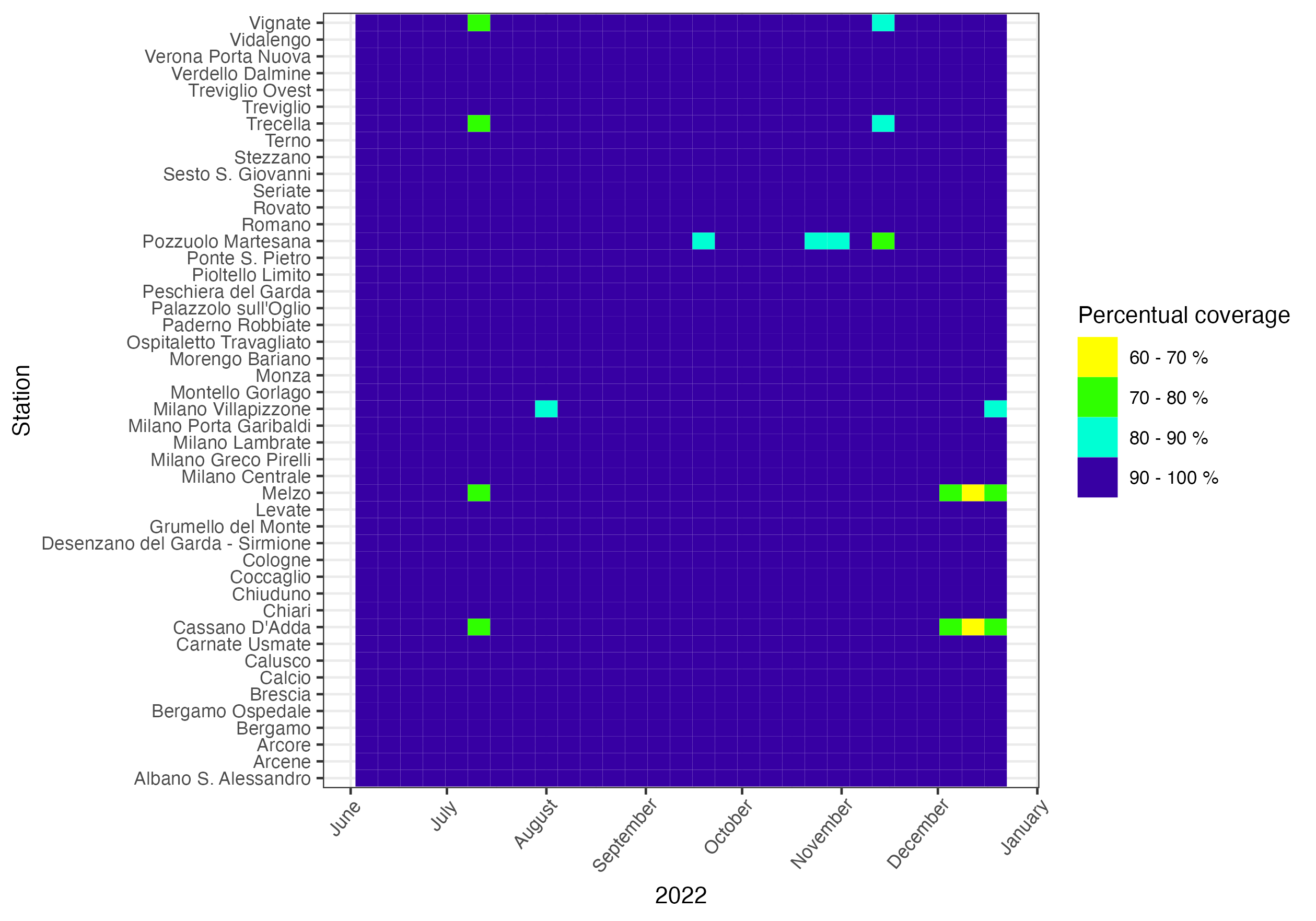}
    \caption{Coverage $coverage_i^{[w]}$ per week $w \in W$ and station $i \in S$ during seven months of 2022.}
    \label{fig:coverage}
\end{figure}

\Cref{fig:coverage} presents the distribution of $coverage_i^{[w]}$ across station-week pairs, showcasing that most station-week pairs have coverage exceeding 90\%. Notably, even the lowest value observed in the dataset is 67\%, indicating substantial coverage of train rides during the study period. This motivates us in applying~\Cref{eq:Rescaling_trains} to estimate the number of boarded and alighted passengers during each week, as missing data are a minority in the counter dataset.

\begin{figure}[htb]
    \centering
    \includegraphics[width=\textwidth]{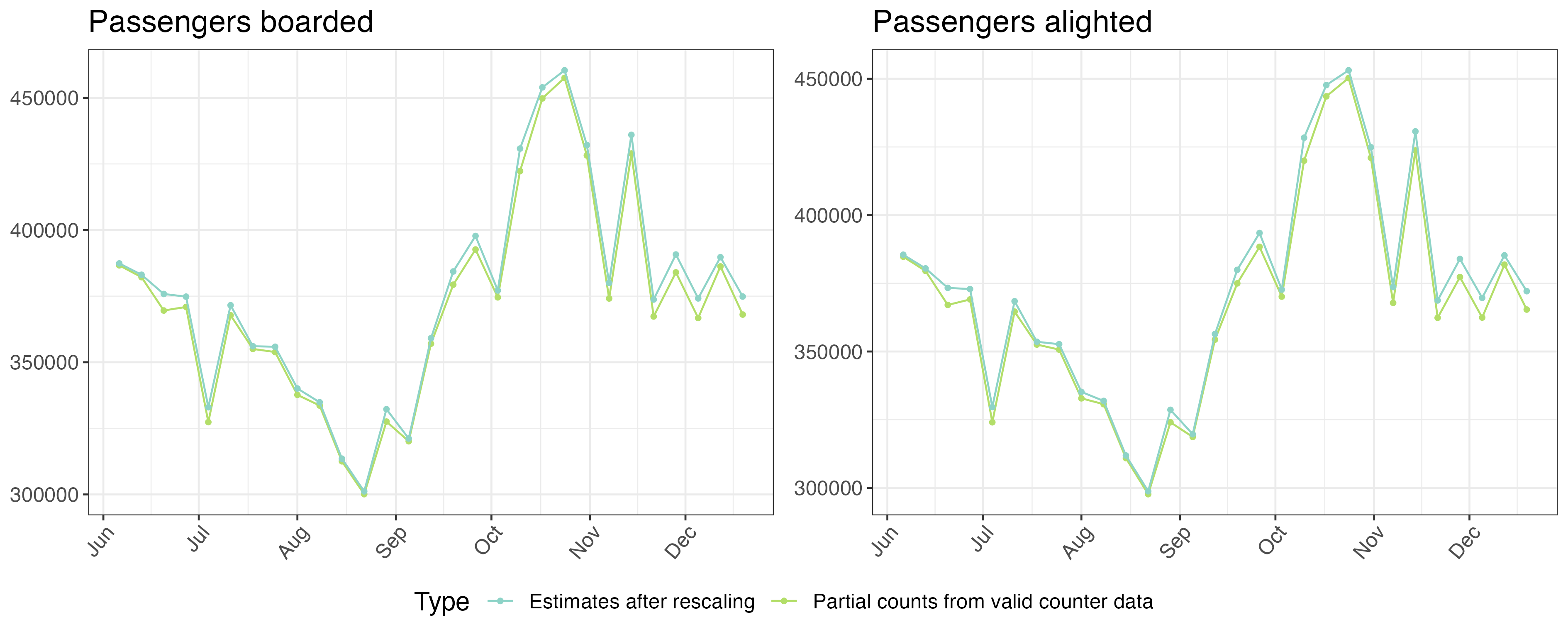}
    \caption{Weekly counts of boarded passengers (left) and alighted passengers (right) from June to December 2022. The light-green line represents partial passenger counts considering train rides with valid estimates from the APC system, while the dark-green line depicts estimates for passenger counts applying the rescaling in~\Cref{eq:Rescaling_trains}.}
    \label{fig:counter_total}
\end{figure}

\Cref{fig:counter_total} shows the estimated number of passengers who boarded and alighted a train during the study period, based on partial valid counter data and following the rescaling procedure detailed in~\Cref{sec:pipeline_counter} using~\Cref{eq:Rescaling_trains}. It is evident that the rescaling has not altered the passenger trends for boarding and alighting. Passenger counts vary throughout the study period, with lower values during the summer and subsequent increases in September, followed by a decrease during the Christmas holidays.

\subsection{Conversion of ticket data into seed Origin-Destination matrices}

Applying the pipeline outlined in~\Cref{sec:pipeline_ticket}, we generate 29 seed OD matrices $X^{*[w]} \ \forall w \in W$ based on ticket data, portraying movements inferred from tickets and subscriptions purchased within the seven months of 2022 considered in our study and the month of May. This process is achieved through three steps, as follows:
\begin{enumerate}
    \item Converting each record within the ticket dataset into one or multiple trips characterized by an origin ($i \in S$), destination ($j \in S$), and week ($w \in W$).
    \item Separating trips necessitating transfers according to the optimal interchange station derived from timetable data.
    \item Estimating missing ticket data using a gravity model.
\end{enumerate}

\begin{figure}[ht]

    \begin{subfigure}{.5\textwidth} % this sets the figure to be max half the width of the page
        \centering
        % include first image
        \includegraphics[width=.9\linewidth]{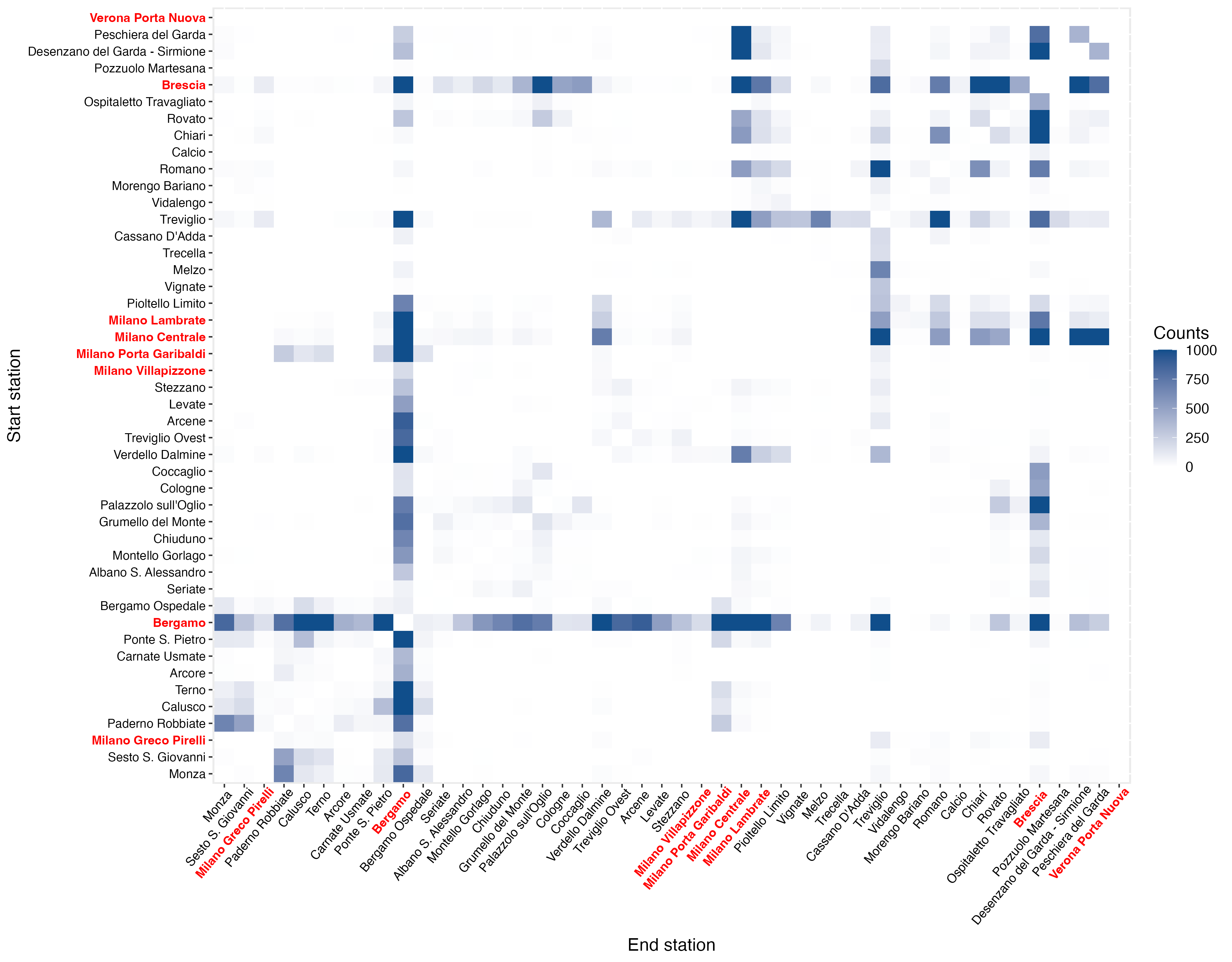}  % this sets the image to fill 90% of the available space -> 45% of the line width in total. 
        \caption{OD matrix after ticket-to-trip conversion.}
        \label{fig:OD_ticket_1}
    \end{subfigure}
    \begin{subfigure}{.5\textwidth}
        \centering
        % include second image
        \includegraphics[width=.9\linewidth]{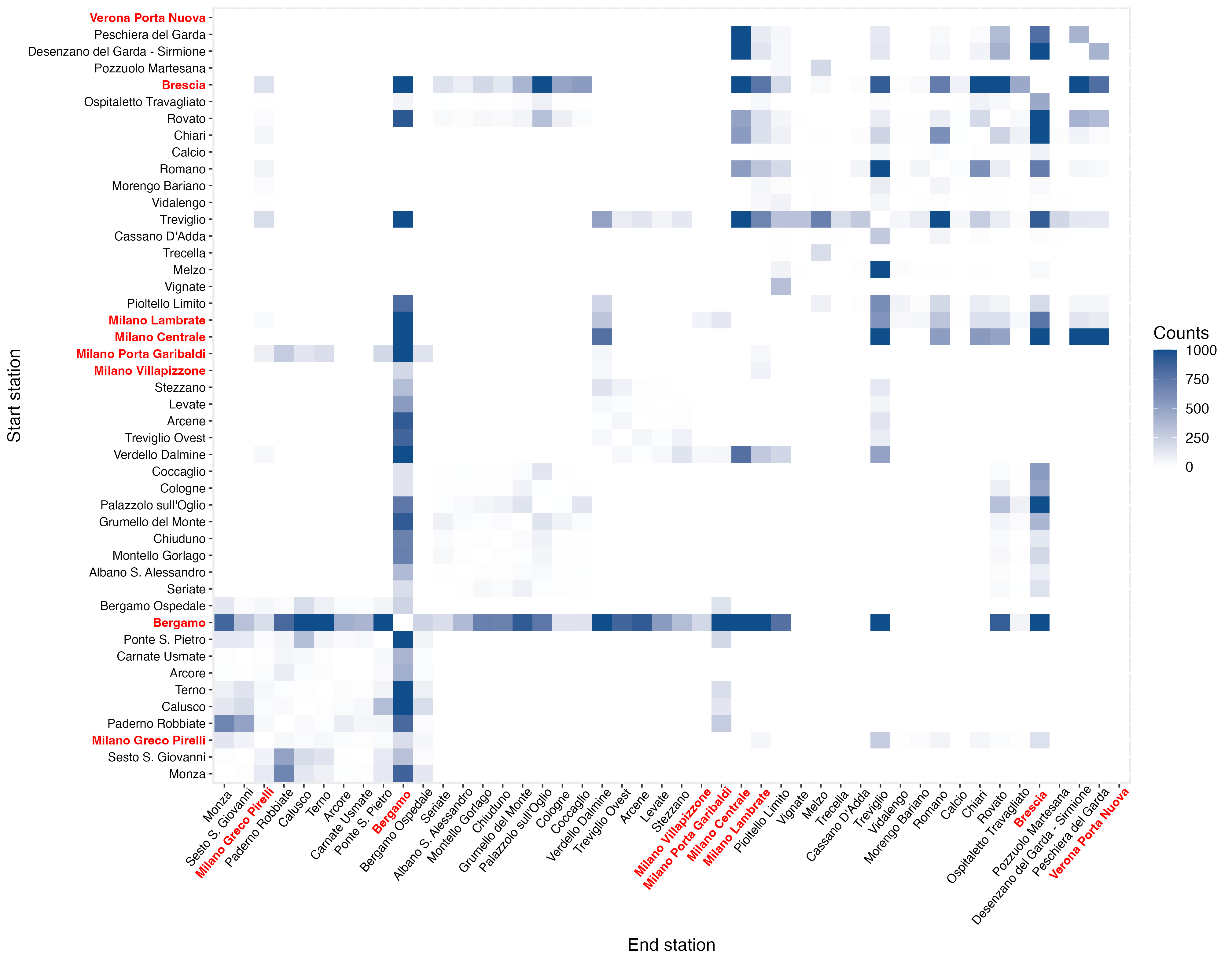}  
        \caption{OD matrix after separating trips requiring transfers.}
        \label{fig:OD_ticket_2}
    \end{subfigure}

    % Leave some vertical space
    \vspace{10pt}

    \begin{subfigure}{\textwidth}
        \centering
        % include first image
        \includegraphics[width=.45\linewidth]{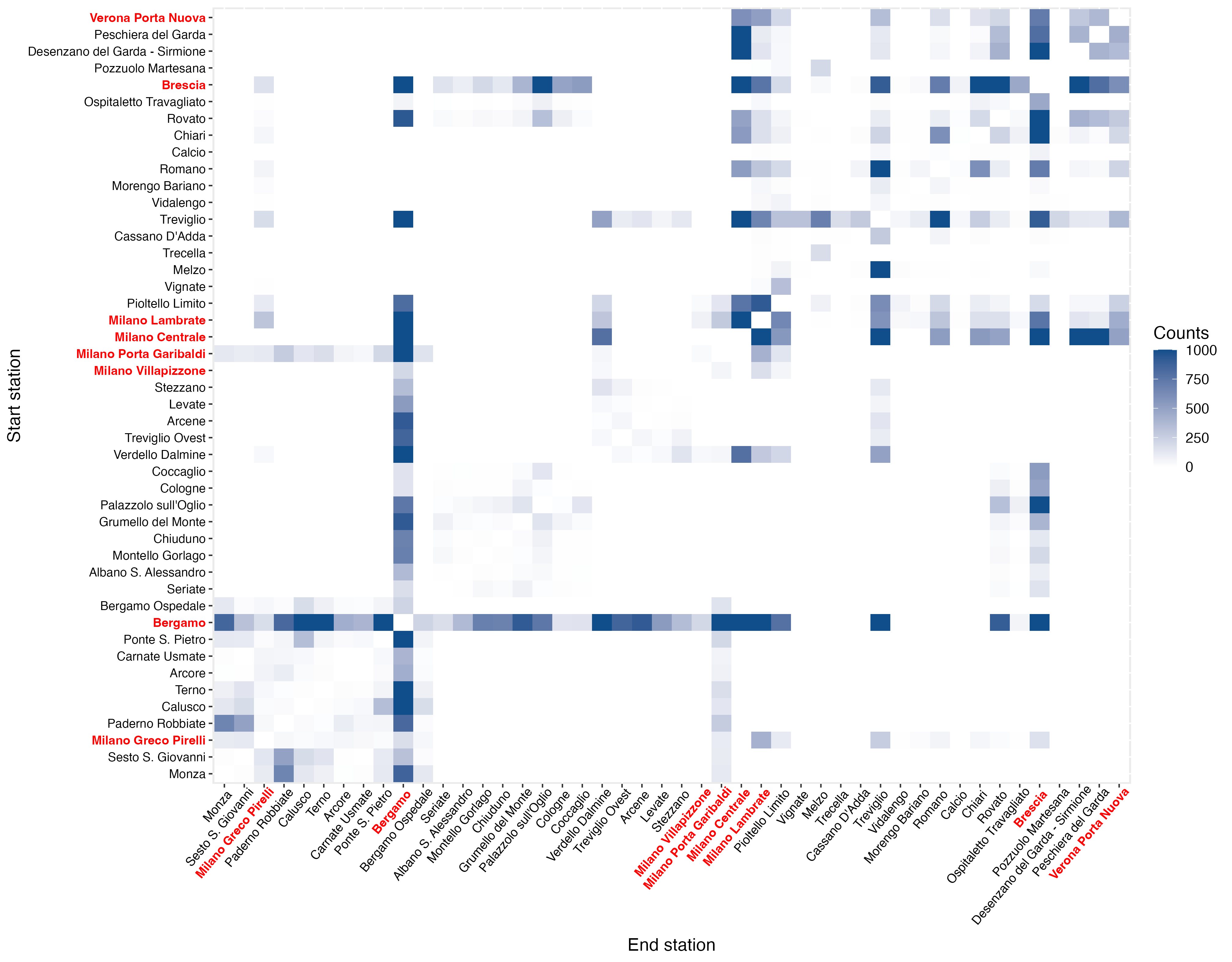}   % this width should be half of the width of the other two images
        \caption{OD matrix after estimating missing ticket data.}
        \label{fig:OD_ticket_3}
    \end{subfigure}

    \caption{Ticket-estimated OD matrix $X^{[38]*}$ following the three-step procedure for generating OD seeds. This example pertains to the week starting on September 19, 2022, and ending on September 25, 2022. The main stations of the network are highlighted in red.}
    \label{fig:OD_ticket}
\end{figure}

\Cref{fig:OD_ticket} illustrates this procedure for the example of the 38\textsuperscript{th}  week of the year (September 19-25, 2022), showcasing the evolving ticket-estimated OD matrix ($X^{[38]*}$) after each step.

In detail,~\Cref{fig:OD_ticket_1} displays the ticket-estimated OD matrix achieved by applying the ticket-to-trip conversion assumptions. It's noteworthy that there are no estimated trips to or from \textit{Verona Porta Nuova} station and no trips between internal Milan stations (\textit{Milano Lambrate, Milano Centrale, Milano Porta Garibaldi, Milano Villapizzone, Milano Greco Pirelli}) and stations within the IS area (\textit{Monza, Sesto S. Giovanni, Arcore, Carnate Usmate, Pioltello Limito, Vignate, Melzo, Trecella, Cassano d'Adda}, and \textit{Pozzuolo Martesana}). The absence of data in these cells is due to missing ticket data for these paths.

Conspicuous movements are detected around stations Brescia, Treviglio, and Bergamo in all the matrices, reflecting their high activity levels within the network. Furthermore, a decrease is observed in all cells during August, followed by an upswing in early September after the end of the summer period.

After constructing the ticket-estimated OD matrices, the subsequent task is to separate trips requiring a transfer based on the optimal interchange station determined in~\Cref{sec:pipeline_times}. This process omits paths necessitating more than one station change, constituting a mere 0.7\% of the total estimated trips from ticket data, as shown in~\Cref{table:tickets_transfers}. Consequently, only a negligible fraction of data is lost by excluding trips with multiple transfers.

\begin{table}[htb]
    \centering
    \begin{tabular}{lll}
      \toprule
      Transfers' number & Estimated trips & \% of total estimated trips \\
      \midrule
      Direct paths & 4,770,557 & 95.2\% \\
      One transfer & 235,316 & 4.70\% \\ 
      More than one transfer & 3,822 & 0.76\% \\
      \bottomrule
    \end{tabular}
    \\[10pt]
    \caption{Number of estimated trips and percentual of total estimated trips across seven months of 2022, divided into direct paths, one transfer and more than one transfer. }
    \label{table:tickets_transfers}
  \end{table}

\Cref{fig:OD_ticket_2} exhibits the same ticket-estimated OD matrix after separating trips requiring transfers. Notably, only OD paths directly connected in $\mathcal{D}$ exhibit non-zero cells in $X^{[w]*}$.

The final step in ticket data processing is estimating missing OD cells, encompassing paths to and from \textit{Verona Porta Nuova} station and routes linking Milan stations with other IS area stations. This estimation employs the gravity model described in~\Cref{eq:gravity_model_Trenord}. The estimates of coefficients, p-values signifying parameter significance, and the $R^2$ coefficient for the gravity model are provided in~\Cref{table:gravity_model_estimates}. As it is natural to expect in the formulation of the gravity model, the coefficients $\alpha$ and $\beta$ which express the influence of the number of passengers boarding and alighting at the origin and destination station on the OD flow are positive while $\gamma$, representing the effect of the separation in time between the two stations, is negative.

\begin{table}[htb]
    \centering
    \begin{tabular}{lll}
      \toprule
      Variable     & Estimate & p-value   \\
      \midrule
      (Intercept) & -8.67 & (.000) \\ 
    $\alpha$ & 0.68 & (.000) \\ 
    $\beta$ & 0.92 & (.000) \\
    $\gamma$ & -0.35 & (.000) \\
    \midrule
    $R^2$ & 0.56 & \\ 
      \bottomrule
    \end{tabular}
    \\[10pt]
    \caption{Coefficient estimates for the gravity model presented in~\Cref{eq:gravity_model_Trenord}, along with p-values denoting parameter significance.}
    \label{table:gravity_model_estimates}
\end{table}

This model is employed to predict the 70 cells related to paths with missing ticket data for every week $w \in W$. \Cref{fig:OD_ticket_3} displays matrix $X^{[38]*}$ again, following the prediction of missing ticket data. Notably, cells linked to \textit{Verona Porta Nuova station} and paths between Milan and IS area stations now contain non-zero values.

After the steps taken to obtain seed OD matrices from ticket data, which generates estimates for railway trip movements, these matrices serve as input for the IPF method. The algorithm corrects them by adjusting each cell with suitable values to ensure their rows and columns sum to the count of boarded and alighted passengers.

\subsection{Application of the Iterative Proportional Fitting algorithm}
Following the conversion of tickets into seed OD matrices ($X^{*[w]}$) and the computation of the margin vectors ($p_i^{[w]}$ and $a_i^{[w]}$) representing the total number of passengers boarded and alighted at each station $i \in S$ during week $w \in W$, we can now proceed to generate the estimated OD matrices ($X^{[w]}$) describing train movements from station $i$ to station $j$ in week $w$ by applying the IPF algorithm as specified in~\Cref{sec:pipeline_IPF}.

\begin{figure}[htb]
    \centering
    \includegraphics[width=.6\textwidth]{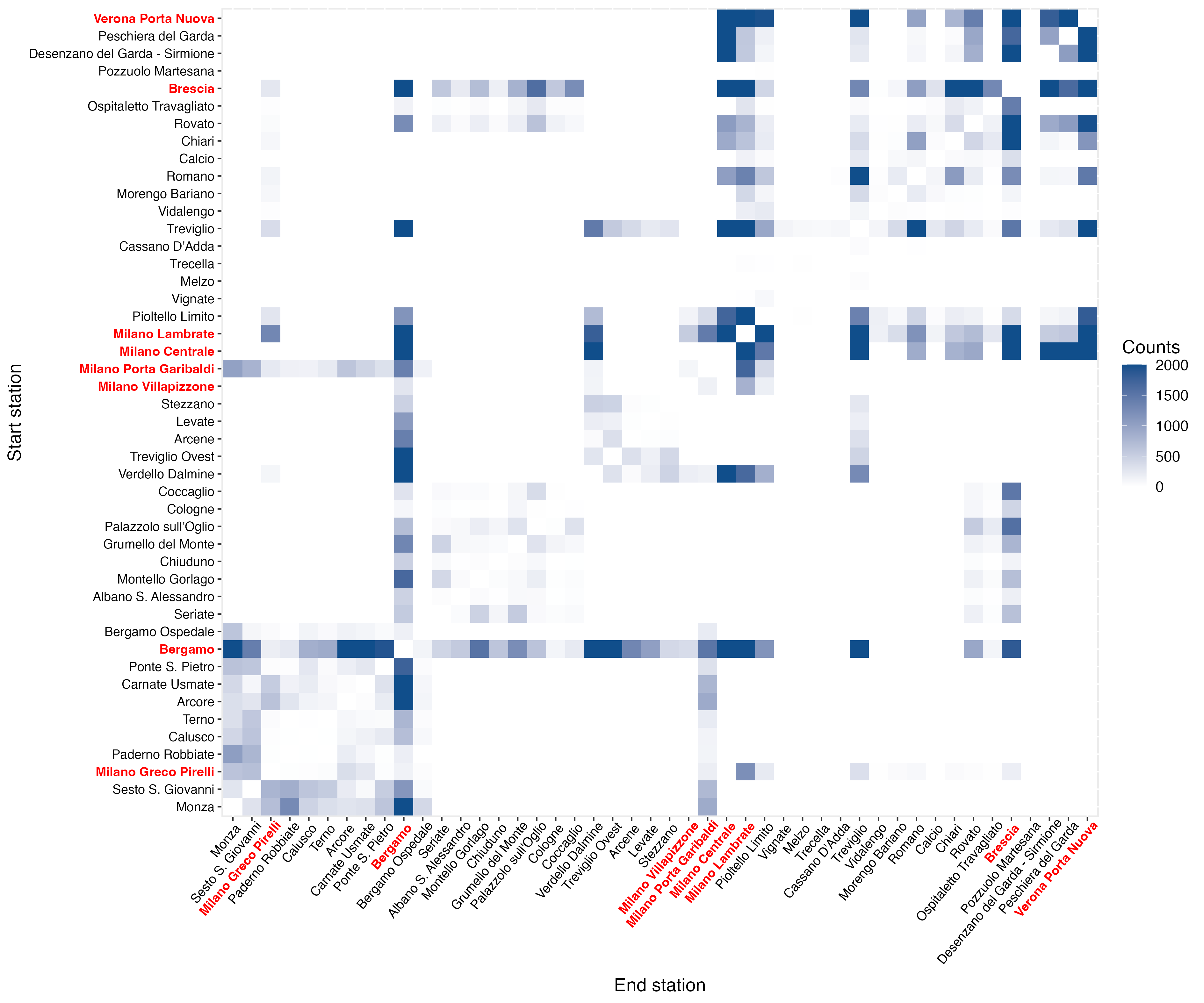}
    \caption{OD matrix $X^{[38]}$ estimated using the IPF method, illustrating train movements in the railway network for the week starting on September 19, 2022, and ending on September 25, 2022. The main stations of the network are highlighted in red.}
    \label{fig:Furness_38}
\end{figure}

An example of one of the finalized OD matrices is presented in~\Cref{fig:Furness_38}. This matrix corresponds to the same week as the ticket-estimated seed OD matrix shown in~\Cref{fig:OD_ticket}.

The final OD matrices exhibit adherence to real-world principles:
\begin{enumerate}
    \item Noticeable movement centers around major hubs such as Bergamo, Brescia, Verona and Milan stations.
    \item Stations linked by two or three lines display greater movement than stations with only a single connection.
    \item Mobility experiences a decline during the summer, notably in August.
    \item Mobility sees a reprise around the start of September.
    \item Mobility declines once more during the Christmas season.
\end{enumerate} 

\begin{figure}[htb]
    \centering
    \includegraphics[width=0.3\textwidth]{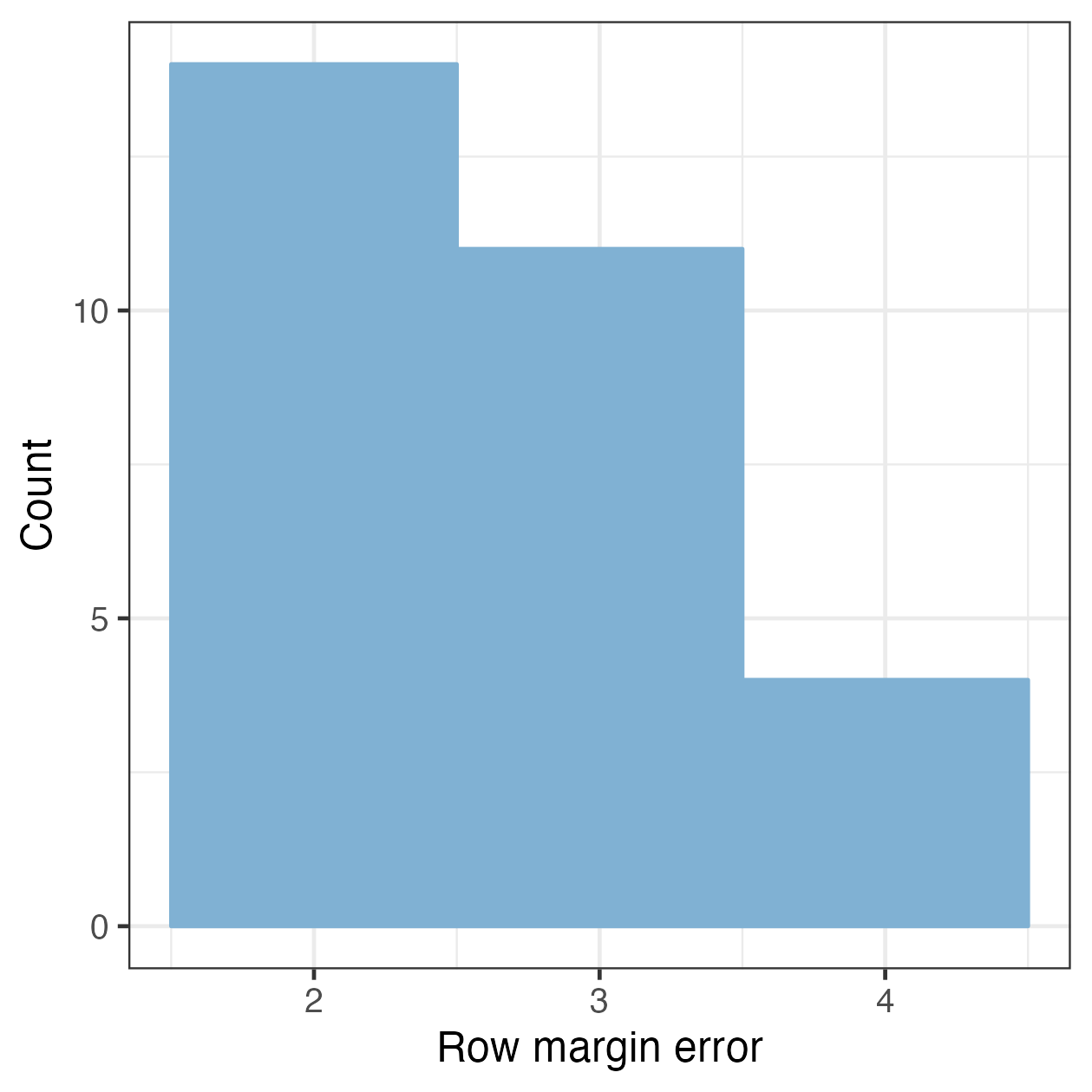}
    \caption{Histogram of row margin errors $\epsilon_{row}^{[w]}$.}
    \label{fig:margins_errors}
\end{figure}

Furthermore, we can quantify the alignment between the final matrix $x_{ij}^{[w]}$ and the margins $p_i^{[w]}$ and $a_j^{[w]}$, assessing margins errors $\epsilon_{row}$ and $\epsilon_{col}$ as the maximal deviation between each calculated and desired margin, as detailed in~\Cref{eq:Furness_errors}.
It is important to note that since the final matrix results from the product of the probability matrix $\pi_{ij}^{[w]}$ and the total boarded passengers, as in~\Cref{eq:Furness_probability}, we should consider row errors $\epsilon_{row}^{[w]}$ to judge the fitting quality of the procedure. The row margin errors $\epsilon_{row}^{[w]}$ are consistently low, as shown in~\Cref{fig:margins_errors}. This strongly indicates that the margins of the estimated cells $x_{ij}^{[w]}$ align well with the actual margins $p_i^{[w]}$, as expected.

\subsection{Exploiting Origin-Destination matrices}

Dynamic OD matrices offer several potential usages in sustainable planning as well as in complex systems maintenance, for example, anomaly detection within the Trenord network: dynamic OD matrices may be employed in global and local identification of abrupt changes in mobility expressed by the network revealing insights into the network's behavior. To achieve this, we leverage network analysis techniques~\cite{NetworkAnalysis}, treating dynamic OD matrices as weighted directed dynamic networks. This, coupled with tools deriving from functional data analysis (FDA)~\cite{FDA}, allows for the maintenance of a time series of meaningful indicators of mobility patterns at both global and local levels.

\subsubsection{Global indicators}
Global indicators analyze the network characteristics and showcase the differences between consecutive weeks. We defined two global indicators and interpreted them together with some events that may have influenced the network's dynamics. These events are divided into strikes, holidays, and infrastructural construction work. \Cref{appendix:AppendixB} reports a description of all the events identified in the period from June to December 2022. 

\begin{figure}[htb]
    \centering
    \includegraphics[width=0.8\textwidth]{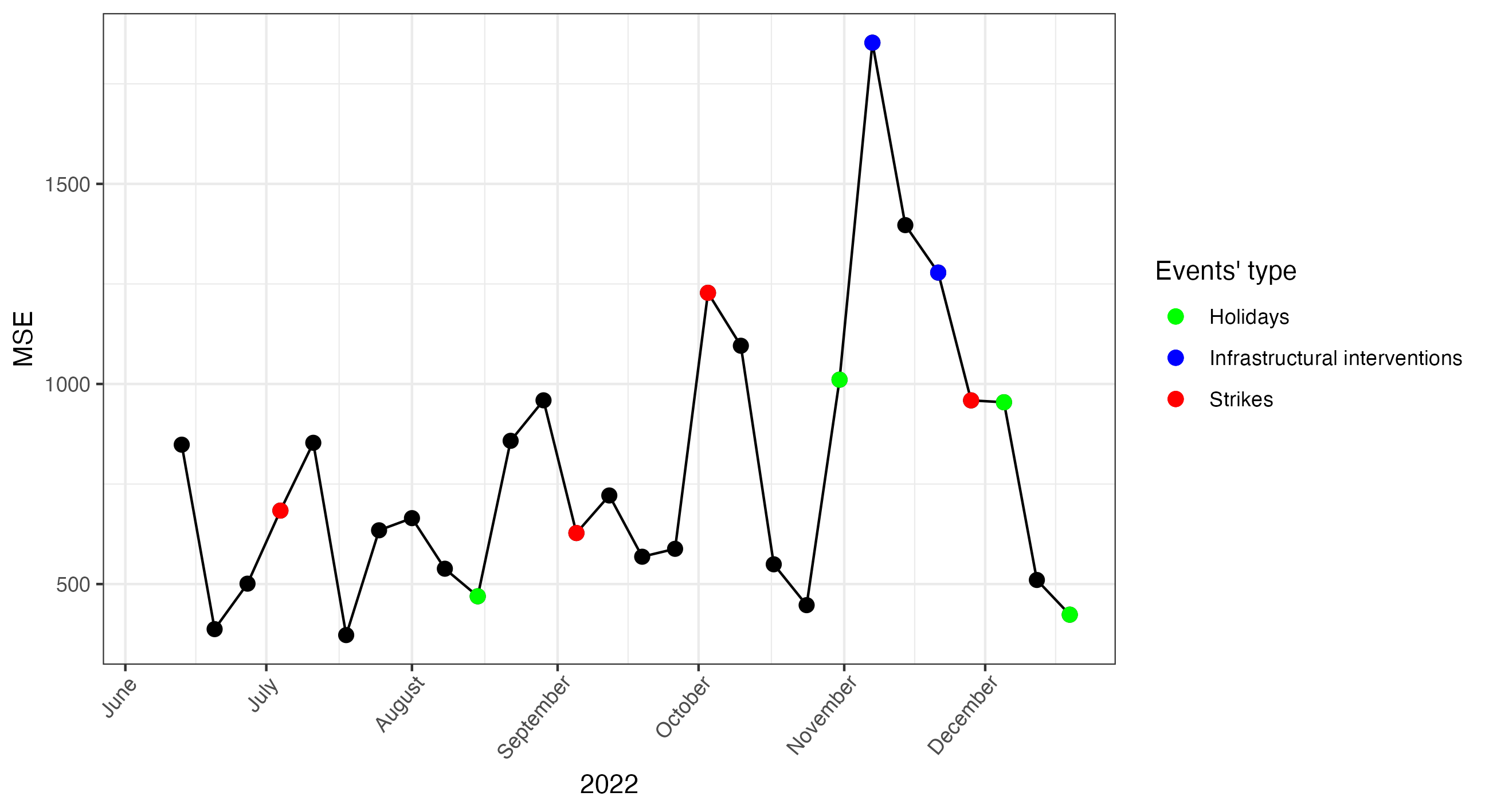}
    \caption{Values of $MSE^{[w]}$ for each of the 29 weeks of the study. Colored points indicate weeks when one of the events reported in~\Cref{appendix:AppendixB} has happened.}
    \label{fig:MSE}
\end{figure}

The global indicators we defined are:

\paragraph{Mean squared error (MSE)} This indicator, defined as
\begin{equation*}
    MSE^{[w]} = \frac{1}{S^2} \sum_{i=1}^{S} \sum_{j=1}^{S} (x_{ij}^{[w]} - x_{ij}^{[w-1]})^2
\end{equation*}

measures differences between subsequent weeks in the network. High $MSE^{[w]}$ values mean significant global fluctuations in movements between the week $w$ and the previous one $w-1$, while low values indicate stability between subsequent weeks. Events such as strikes, holidays, and infrastructural interventions, reported in~\Cref{appendix:AppendixB}, were noted as potential influencers. For instance, the two major network interventions resulted in notable disruptions, leading to spikes in MSE values. These observations are depicted in~\Cref{fig:MSE}.
 
\paragraph{Mean strength} As a second global indicator, we compute the mean strength as the average number of passengers passing through the network's stations as:
\begin{align}
    \label{eq:strength}
    \sigma_i^{[w]} &= \sum_{j=1}^{S} x_{ij}^{[w]} + \sum_{j=1}^{S} x_{ji}^{[w]} \\
    \bar\sigma^{[w]} &= \frac{1}{S} \sum_{i=1}^{S} \sigma_i^{[w]}
\end{align}

Notice that the strength $\sigma_i^{[w]}$ could be equivalently computed directly from the number of boarded and alighted passengers as: 
\begin{equation*}
    \sigma_i^{[w]} = p_i^{[w]} + a_i^{[w]}
\end{equation*}

\begin{figure}[htb]
    \centering
    \includegraphics[width=0.8\textwidth]{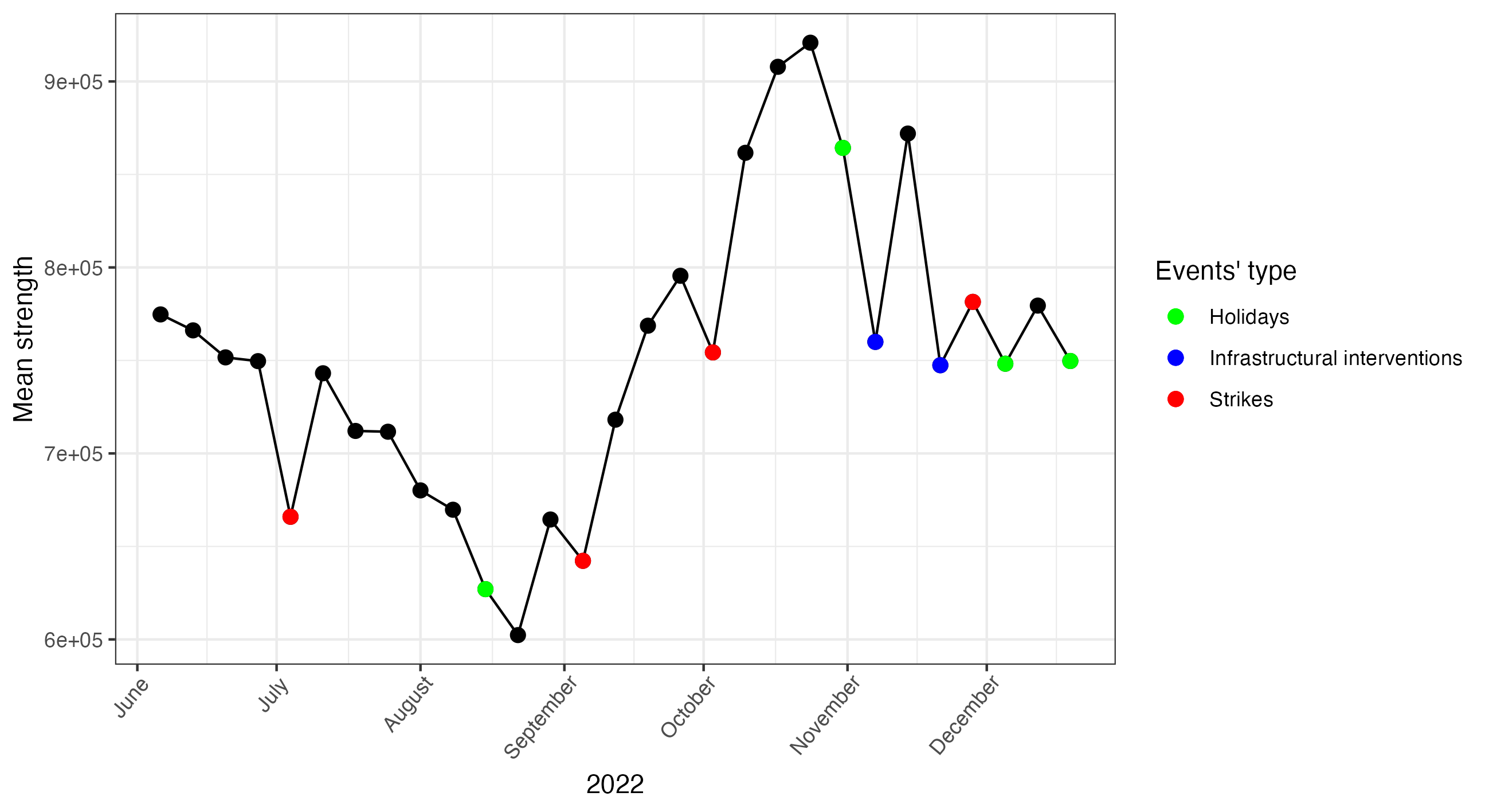}
    \caption{Values of $\bar\sigma^{[w]}$ for each of the 29 weeks of the study. Colored points indicate weeks when one of the events reported in~\Cref{appendix:AppendixB} has happened.}
    \label{fig:mean_strength}
\end{figure}

This indicator is again influenced by events and holidays, leading to fluctuations in mobility trends. Notably, a dip in mean strength is observed in August when many Italian companies typically close for summer holidays. \Cref{fig:mean_strength} illustrates these fluctuations and their correlation with events.
  
Thus, the two global indicators defined, $MSE^{[w]}$ expressing variations in mobility in subsequent weeks and $\bar\sigma^{[w]}$ depicting the network's global mobility for each week, are able to identify weeks where the network's disruptions can be traced back to some events affecting railway mobility. 

\subsubsection{Local indicators}
After identifying weeks of anomalous mobility behavior, we turn to the local station level to study the evolution of network indicators and assess how each node reacts to the network's disruptions. We consider the weekly strength $\sigma_i^{[w]}$ for each station $i \in S$, as defined in~\Cref{eq:strength}. To remove the effect of the volume of passengers passing through each node and analyze only mobility trends, we consider the normalized version of the strength by dividing each strength value by the sum of the station's strengths through the 29 weeks of the study as:
\begin{equation}
    \widetilde{\sigma}_i^{[w]} = \frac{\sigma_i^{[w]}}{\sum_{w \in W} \sigma_i^{[w]}}
\end{equation}

\begin{figure}[htb]
    \centering
    \includegraphics[width=.8\textwidth]{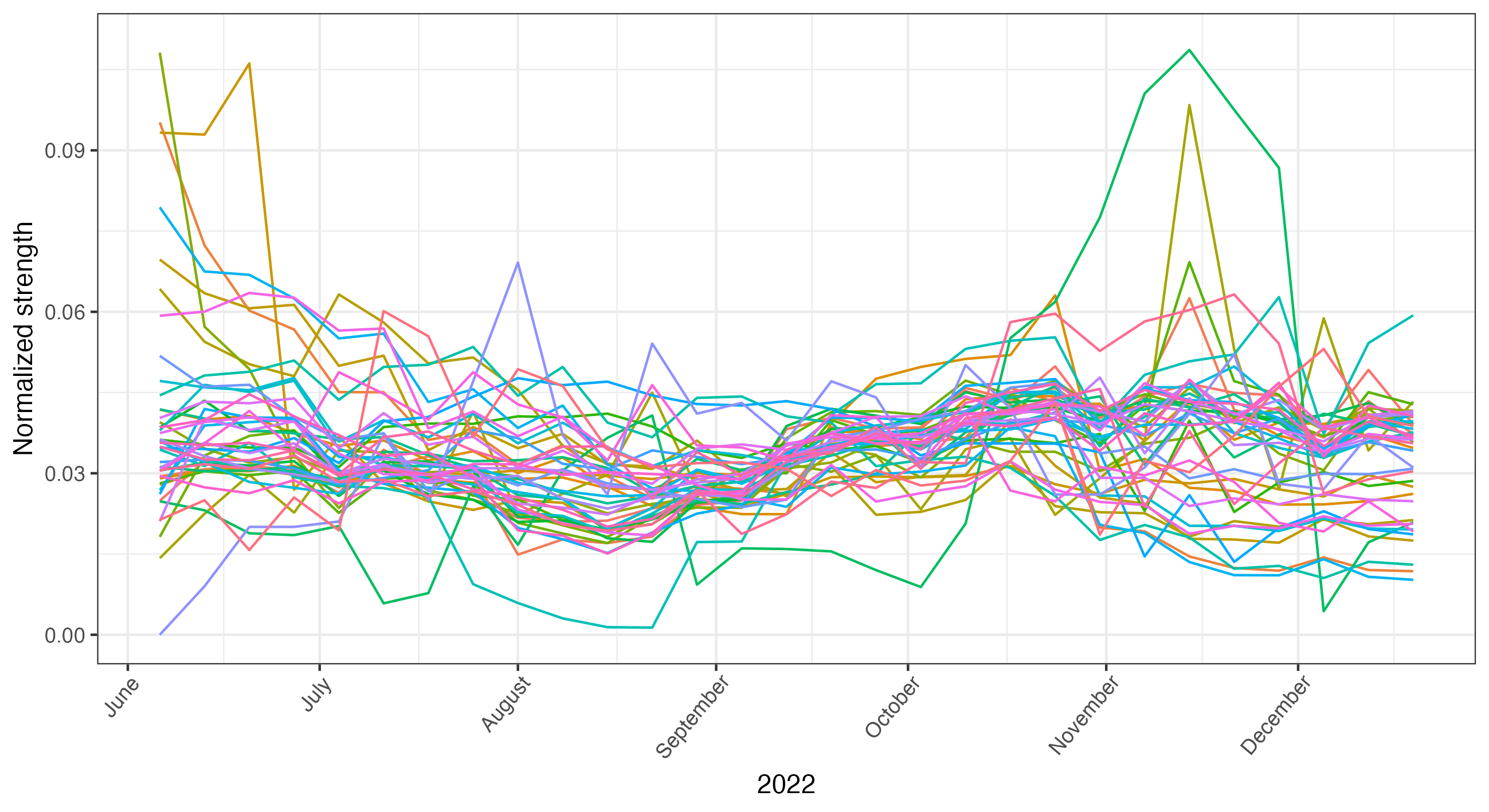}
    \caption{Values of $\widetilde{\sigma}_i^{[w]}$ for each of the 46 stations $i \in S$ of the study.}
    \label{fig:strength_normalized_stations}
\end{figure}

\Cref{fig:strength_normalized_stations} reports the values of $\widetilde{\sigma}_i^{[w]}$ for the 46 stations of the network. The figure allows us to notice the heterogeneity in normalized mobility trends between the stations through the seven months of the study.
 
\begin{figure}[htb]
    \centering
    \includegraphics[width=.8\textwidth]{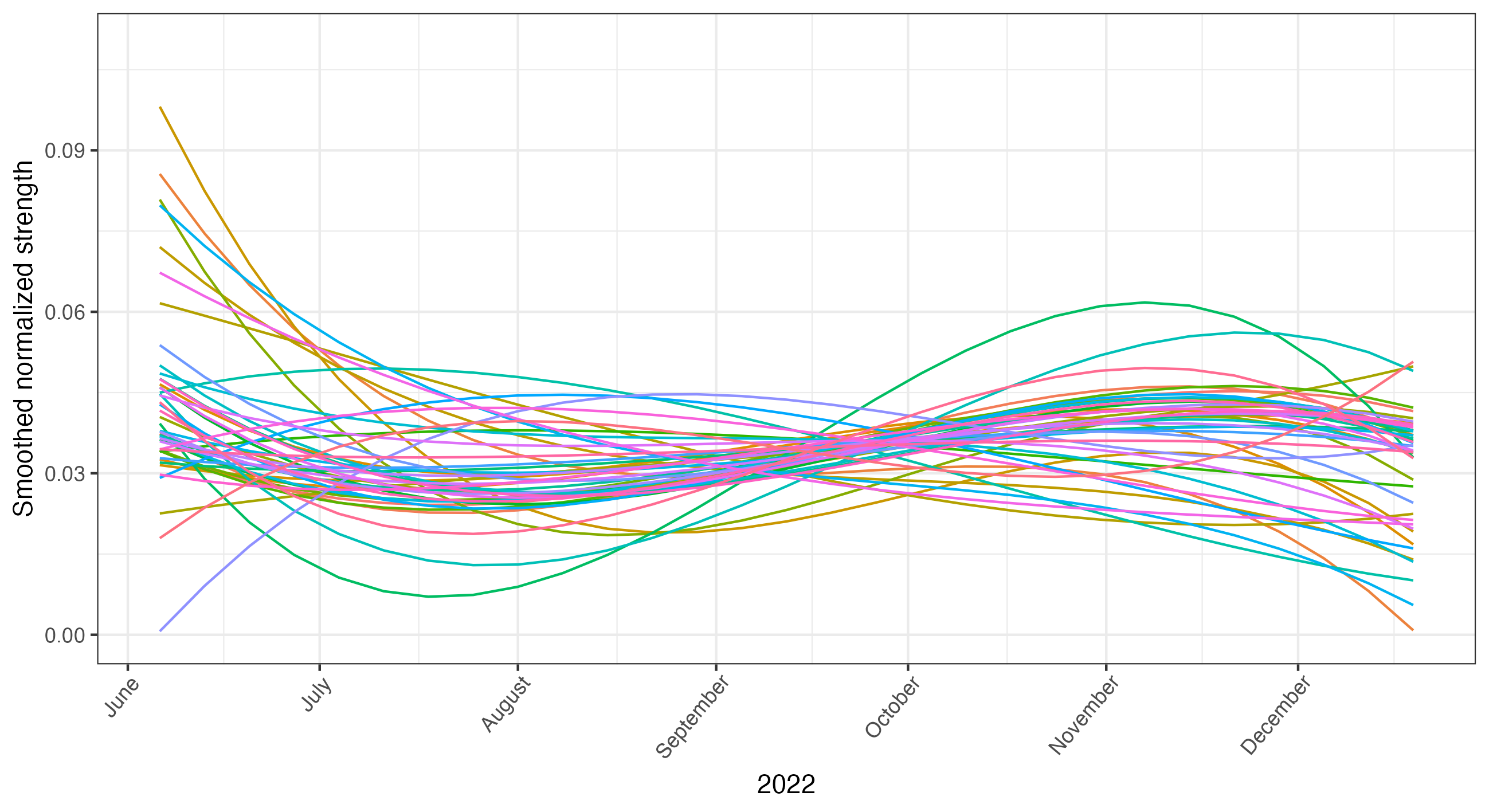}
    \caption{Smoothed function of $\widetilde{\sigma}_i^{[w]}$ for each of the 46 stations $i \in S$ of the study.}
    \label{fig:strength_normalized_smoothed}
\end{figure}

To identify anomalous mobility trends at the station level, we apply FDA techniques. In particular, we compute the smoothed mobility density of $\widetilde{\sigma}_i^{[w]}$ through cubic splines basis, using four basis functions and adding a roughness penalty on the second derivative of the curves. We selected the number of basis functions minimizing the average generalized cross-validation error across all the curves. \Cref{fig:strength_normalized_smoothed} shows the 46 smoothed functions during the 29 weeks of the study. 

\begin{figure}[htb]
    \centering
    \includegraphics[width=.8\textwidth]{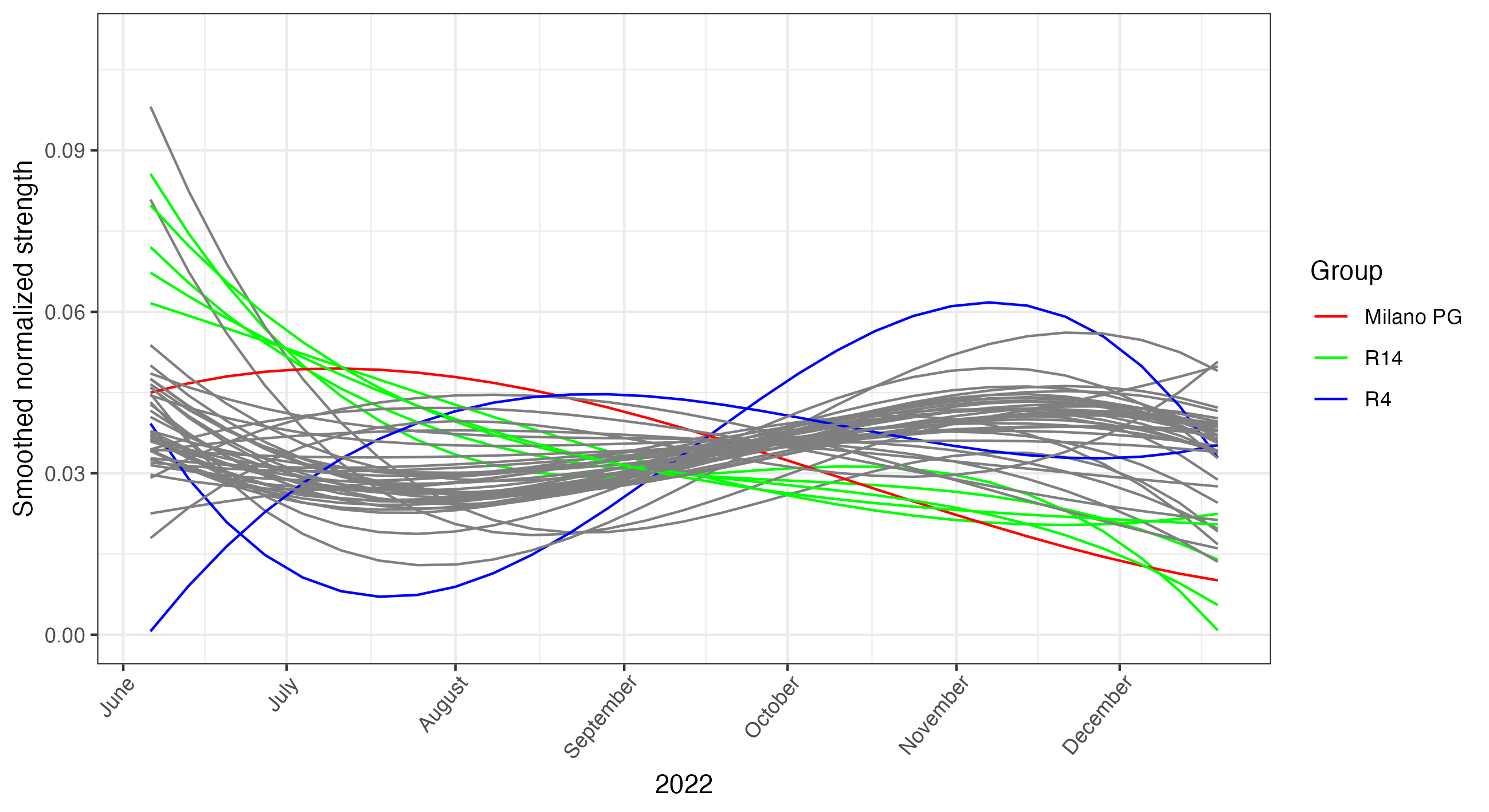}
    \caption{Functional outliers revealed by the functional boxplot, interpreted together with the train lines whose station belongs to.}
    \label{fig:functional_outliers}
\end{figure}

The smoothed functions are the starting point for further analyses involving FDA techniques, such as but not limited to functional clustering, outlier detection and principal components analysis. Since this Section focuses on highlighting anomalous mobility trends, we evaluate outliers based on the functional boxplot~\cite{FunctionalBoxplot}, which revealed 8 of the 46 stations as potential outliers. These 8 stations are shown in \Cref{fig:functional_outliers}, categorized by their train lines: stations on line R14 display less heterogeneity in oscillations than the others, while those on line R4 exhibit higher amplitudes, indicating higher variability. Additionally, station \textit{Milano Porta Garibaldi} shows oscillations in phase opposition to the average behavior. 

Thus, this Section demonstrated various tools to identify anomalies in the Trenord network using dynamic OD matrices. These insights highlight the utility of such matrices in understanding network behavior, offering valuable applications for transportation operators and policymakers.

\section{Discussion}
\label{sec:discussion}

This work has presented an innovative pipeline centered around the IPF algorithm to address the challenge of estimating dynamic OD matrices within a railway transportation network. We now highlight the strengths and criticalities of our work, as well as potential avenues for future research.

Our pipeline is designed to estimate weekly OD matrices for a railway network by combining data from ticket and subscription sales with passenger counts collected through the APC system. While recent studies have emerged in the field of OD matrix estimation in public transport networks through ACDS systems, they often rely on AFC data, particularly smart card information, and travel surveys. In contrast, our approach demonstrates a methodology to derive accurate OD matrices without necessarily having access to such data. We introduce the concept of constructing seed OD matrices using ticket and subscription sales data, assuming they are available in several transportation companies. Thus, the primary innovation of our research lies in the procedure developed to transform ticket data into OD seeds. The IPF method, a well-established technique for combining OD seeds with passenger count data to refine OD matrices, is then applied to fuse the ticket-estimated OD seeds with the APC-derived passenger counts. Our approach can be expanded to encompass more extensive transportation networks or different modes of transportation, such as metro systems, bus networks, or other train services, provided we have sufficient data available to derive OD seeds and passenger counts.

The estimation of passenger counts within our pipeline relies on APC data, which are expected to be installed on an increasingly growing proportion of transportation systems~\cite{Siebert2020}, making passenger counts readily available without the need for estimating unknown counts.
However, the actual contribution of our work within the realm of trip distribution modeling stands in the processes we designed to generate OD seeds from ticket and subscription sales data. This step, essential for obtaining reliable seed OD matrices critical for applying the IPF algorithm~\cite{IPFP_problems}, encapsulates  the main novelty of our study. While numerous studies focus on OD matrix estimation, to the best of our knowledge, none have delved into the challenge of deriving accurate OD seeds from ticket and subscription sales data, fusing such data with passenger counts. 
Subsequently, the counter-derived marginal data, representing boarded and alighted passengers at each station and week, is combined with the ticket-estimated OD seeds using the IPF algorithm. This iterative approach generates 29 weekly OD matrices, portraying train movements across the six lines within the Trenord network considered in our study. These matrices adhere to real-world principles, depicting high movement between stations connected by multiple lines and reduced trips during the summer and Christmas periods. The low row margin errors, indicating the agreement between the estimated matrices and margin vectors, coupled with the consistent convergence of the iterative process, lend confidence to the robustness of our results. Moreover, we compared the dynamic OD matrices with a static OD matrix representing mobility in the Lombardy area in 2020~\cite{RL_OD_data}. We found a notable correlation between the two datasets, validating the pipeline’s capacity to yield dynamic OD matrices approximating actual mobility. 

The pipeline we developed is fast and can be readily applied at each timeframe needed as soon as data becomes available, leading to the possibility of obtaining dynamic OD matrices describing the more recent developments in the network. Thus, the estimation of dynamic OD matrices serves as the foundation for analyses of the network's behavior relying on accurate, dynamic and recent mobility data. For example, such matrices can enhance operational insights for transportation operators, supporting demand study and schedule optimization to match fluctuating demands throughout the year. Furthermore, by applying our methodology to various temporal periods, such as pre and post-COVID-19 pandemic times, these matrices offer valuable tools for exploring lifestyle changes and assessing shifts in commuting patterns. These matrices can also facilitate environmental impact assessments for railways, enabling calculations of carbon emissions, fuel consumption, and pollution levels, serving as a basis for comparisons against other transportation modes. 

In the framework of possible usage of dynamic matrices, we showed an example of such applications, proving that the derived dynamic OD matrices offer valuable insights into the network’s mobility patterns and disruptions. We focused on anomaly detection, both at the global network and local station level, representing the dynamic OD matrices as temporal weighted networks. We applied a set of network and functional data analysis techniques to highlight that the derived OD matrices offer valuable insights into the network’s mobility patterns and disruptions. Indeed, we were able to identify anomalies at the global level and match disruptions in the global indicators to events such as strikes, holidays and network interventions. At the local station level, we identified stations showing anomalous mobility trends compared to the others in the period from June to December 2022. 

Looking ahead, assuming that future research efforts will lead toward obtaining accurate dynamic OD matrices in various transportation networks, there is a crucial need for the development of advanced techniques to analyze complex temporal networks induced by dynamic OD matrices. Such analyses could offer valuable insights for policymaking, operational optimization, and urban planning. Accurate dynamic OD matrices provide a real-time snapshot of passenger movements, allowing for a deeper understanding of commuting patterns, peak travel times, and network congestion. By coupling these matrices with sophisticated data analysis methods, researchers can unravel intricate mobility behaviors, offering crucial information for policy formulation and infrastructure planning. Moreover, these dynamic matrices are instrumental in studying the impact of external events, such as pandemics (as was done in~\cite{Galliani2023}), strikes, or major public events, on transportation patterns. By comprehensively understanding these shifts, policymakers can make informed decisions, enhancing the resilience and adaptability of transportation systems to various challenges. Thus, the focus should not only be on accurate data acquisition but also on developing analytical tools capable of harnessing the wealth of information embedded in these dynamic OD matrices. Based on this premise, techniques such as the one developed in~\cite{bail2023flow} may be applied to facilitate the comparative analysis of temporal networks induced by successions of dynamical OD matrices, allowing for the comparison of transportation networks within the same area across different modes of transport or shedding light on networks relating to the same mode but in several geographical locations.

\section{Conclusion}
\label{sec:conclusion} 
This study presented an innovative and comprehensive pipeline for estimating dynamic OD matrices in a railway transportation network. By integrating ticket and subscription sales data with passenger count information collected by the APC system, we have successfully derived accurate weekly OD matrices that depict train movements across six train lines of the Trenord network during seven months of 2022. Our methodology addresses a crucial challenge in transportation network analysis by offering a robust approach for generating OD matrices even in the absence of AFC data or comprehensive survey information, employing ticket and subscription sales, which we expect to be collected and available for most transportation networks.

The cornerstone of our work lies in the development of a novel procedure to construct OD seed matrices from ticket sales data. This entails a multi-step process that involves converting ticket records into OD trips, separating trips requiring transfers, and estimating missing ticket data using gravity models. These steps collectively provide a foundation for reliable OD seed matrices, which are then refined through the IPF algorithm combining such seeds with passenger counts. The resulting dynamic OD matrices offer a comprehensive description of train movements within the Trenord network, capturing fluctuations in demand and mobility patterns over time. The validation of our pipeline's results with an outdated and static OD matrix representing railway mobility in Lombardy in 2020 underscores the effectiveness of our methodology, showing a relevant relationship between the two data sources by applying the methodology presented in~\cite{Galliani2023}.

Beyond its technical contributions, our pipeline carries significant implications for practical applications. Transportation operators and urban planners can exploit the insights from dynamic OD matrices to enhance network optimization and adapt services in response to shifting demands. Furthermore, the scalability of our pipeline enables its real-time implementation, providing up-to-date information on network dynamics and changes in passenger behavior. In this framework, we showed a possible application of dynamic OD matrices relating to anomaly detection of mobility trends at the global and local levels. We identified weeks and stations exhibiting mobility trends different than average behavior and matched them to events connected to network disruptions.

This work opens avenues for future research in the broader context of transportation network analysis. The pipeline's modular design allows for the exploration of alternative methods for different pipeline components and the application to other transportation networks beyond railways. As mobility patterns evolve, incorporating additional data sources could enrich the accuracy and relevance of dynamic OD matrices. Ultimately, our study bridges the gap between ticket sales and passenger count data, offering a comprehensive solution to estimating dynamic OD matrices in complex transportation networks and contributing to a more informed and adaptable approach to urban mobility planning.

\section{Acknowledgments}
We thank Trenord for the collaboration and for sharing the data used in this work. In particular, we thank Dr. Marta Galvani and Dr. Giovanni Chiodi for their support and insightful suggestions. 
The authors acknowledge the support by MUR, grant Dipartimento di Eccellenza 2023-2027. This study was funded by the European Union -  NextGenerationEU, in the framework of the GRINS - Growing Resilient, INclusive and Sustainable project (GRINS PE00000018 – CUP D43C22003110001). The views and opinions expressed are solely those of the authors and do not necessarily reflect those of the European Union, nor can the European Union be held responsible for them.

% Bibliography
\bibliographystyle{unsrt}  
\bibliography{bibliography}  

\appendix
\section{Appendix: Assumptions to convert ticket and subscription sales data into estimated OD trips}
\label{appendix:AppendixA}

\begin{table}[htb]
	\centering
	\begin{tabular}{p{\dimexpr 0.3\linewidth-2\tabcolsep} p{\dimexpr 0.7\linewidth-2\tabcolsep}}
		\toprule
		Ticket type                                                   & Conversion in the dynamic seed OD matrices                                                                                                                                                                                                                                                                                                                                                                                                                                                                                                                        \\
		\midrule
		      
		Ordinary ticket, special rate initiative, additional exaction & We extract a random day in the 7 days following the purchase and attribute 0.5 trips between origin and destination and 0.5 between destination and origin to the extracted day. This is because each ticket can be used in either direction between the two stations for which it has been emitted, so we split the number of trips evenly in the two directions.                                                                                                                                                                                                \\ 
		Carnet                                                        & For carnets, we randomly extract 5 days in the 30 days following the carnet's purchase. We suppose a round trip is made in the 5 days drawn and then aggregate weekly. We chose the period of 30 days to extract the trips because it was previously the validity period of the carnet, while now carnets do not have an expiration date.                                                                                                                                                                                                                         \\ 
		Weekly subscription                                           & For weekly subscriptions, we suppose 5 round trips attributed to the current week if the subscription is bought between Monday and Wednesday, to the following week if the subscription is purchased between Thursday and Sunday.                                                                                                                                                                                                                                                                                                                                 \\
		Monthly subscription                                          & For monthly subscriptions, round trips are distributed into the month's weeks starting from the day of selling. The month of usage is the current month if the subscription is bought before the $22^{nd}$ of the month or the following month if it is purchased on the $22^{nd}$ or the days after. We suppose 5 round trips for full weeks (i.e., entirely belonging to the subscription month). For partial weeks, we use the correspondences obtained by computing and rounding the proportion $\frac{5 \ round \ trips}{7 \ days} * \  n \ partial \ days$. \\
		Yearly subscription                                           & For yearly subscriptions, we attribute 5 round trips to each complete week starting from the day of purchasing and ending the last day of the $12^{th}$ month after purchase, applying the same convention to uncomplete weeks (if any) used for monthly subscriptions                                                                                                                                                                                                                                                                                            \\ 
		\bottomrule
	\end{tabular}
	\\[10pt]
	\caption{Assumptions needed to convert each record in the ticket and subscription sales dataset into estimated OD trips attributed to each week of the study period.}
	\label{table:Trenord_ticket_assumptions}
\end{table}

\section{Appendix: Events influencing the Trenord network's dynamics}
\label{appendix:AppendixB}

\begin{table}[htb]
	\centering
	\begin{tabular}{p{\dimexpr 0.15\linewidth-2\tabcolsep} p{\dimexpr 0.25\linewidth-2\tabcolsep} p{\dimexpr 0.5\linewidth-2\tabcolsep}}
		\toprule
		Event type                                                                          & Date                 & Notes                                                                                                                                                                                  \\
		\midrule
		\multirow[t]{4}{\dimexpr 0.15\linewidth-2\tabcolsep}{Strikes}                       & July 10-11, 2022     & Strike of Trenord's train crews                                                                                                                                                        \\
		                                                                                    & September 9, 2022    & Strike of Trenord's train crews                                                                                                                                                        \\
		                                                                                    & October 8-9, 2022    & Strike of Trenord's train crews                                                                                                                                                        \\
		                                                                                    & December 2, 2022     & Strike of Trenord's train crews                                                                                                                                                        \\
		\midrule
		\multirow[t]{4}{\dimexpr 0.15\linewidth-2\tabcolsep}{Holidays}                      & August 15, 2022      & Assumption Day - National Italian holiday                                                                                                                                              \\
		                                                                                    & November 1, 2022     & All Saints' Day - National Italian holiday                                                                                                                                             \\
		                                                                                    & December 8, 2022     & Immaculate Conception - National Italian holiday. Moreover, December 7, 2022, is Saint Ambrose Day, which is Milan's holiday.                                                          \\
		                                                                                    & December 25, 2022    & Christmas - National Italian holiday                                                                                                                                                   \\
		                                                                                    & December 26, 2022    & St. Stephen's Day - National Italian holiday                                                                                                                                           \\
		\midrule
		\multirow[t]{2}{\dimexpr 0.15\linewidth-2\tabcolsep}{Infrastructural interventions} & November 12-14, 2022 & Infrastructural construction work on line RE\_6, causing the closure of stations \textit{Desenzano del Garda - Sirmione}, \textit{Peschiera del Garda} and \textit{Verona Porta Nuova} \\
		                                                                                    & November 26-28, 2022 & Infrastructural construction work on line RE\_6, causing the closure of stations \textit{Desenzano del Garda - Sirmione}, \textit{Peschiera del Garda} and \textit{Verona Porta Nuova} \\
		
		\bottomrule
	\end{tabular}
	\\[10pt]
	\caption{Events influencing Trenord's network dynamics in 2022.}
	\label{table:events}
\end{table}

\end{document}